\documentclass[twocolumn,preprintnumbers,amsmath,amssymb,floatfix,superscriptaddress]{revtex4-1}

\usepackage{graphicx}
\usepackage{amsmath}
\usepackage{algorithm}
\usepackage{algorithmic}
\usepackage{media9}
\usepackage{epigraph}
\usepackage{smartdiagram}
\usepackage{verbatim}
\usepackage{lmodern}
\usepackage{listings}
\usepackage{tcolorbox}
\usepackage{color,soul}
\usepackage{listings}
\usepackage{minted}
\usepackage{float}

\newcommand{\commentoutA}[1]{}

\begin{document}

\preprint{LA-UR-26-23784}

\title{SEDACS: A Scalable Framework for Complex Chemistry Simulations}

\author{Cheng-Han Li}
\email{chenghanli7777@gmail.com}
\affiliation{Theoretical Division, Los Alamos National Laboratory, Los Alamos, NM 87545, USA}
\author{Joshua Finkelstein}
\affiliation{Theoretical Division, Los Alamos National Laboratory, Los Alamos, NM 87545, USA}
\author{Maksim Kulichenko}
\affiliation{Theoretical Division, Los Alamos National Laboratory, Los Alamos, NM 87545, USA}
\author{Rae A. Corrigan Grove}
\affiliation{Theoretical Division, Los Alamos National Laboratory, Los Alamos, NM 87545, USA}
\author{Sergei Tretiak}
\affiliation{Theoretical Division, Los Alamos National Laboratory, Los Alamos, NM 87545, USA}
\author{Michael E. Wall}
\affiliation{Computing and Artificial Intelligence Division, Los Alamos National Laboratory, Los Alamos, NM 87545, USA} 
\author{Anders M.N. Niklasson}
\email{amn@lanl.gov}
\affiliation{Theoretical Division, Los Alamos National Laboratory, Los Alamos, NM 87545, USA}
\author{Christian F.A. Negre}
\email{cnegre@lanl.gov}
\affiliation{Theoretical Division, Los Alamos National Laboratory, Los Alamos, NM 87545, USA}

\date{\today}

\begin{abstract}
Graph-based linear-scaling electronic-structure theory provides a scalable framework for parallel quantum-mechanical molecular dynamics (QMD) simulations by exploiting the nearsightedness of the non-local electronic connectivity in non-metallic systems. When combined with recent shadow molecular dynamics in an extended-Lagrangian formulation, it enables stable long-time simulations of large, chemically active systems. This article introduces the Scalable Ecosystem, Driver, and Analyzer for Complex Chemistry Simulations (SEDACS), which integrates all these advances within a modular, Python-based software package for large-scale QMD simulations driven by external electronic-structure codes. 
SEDACS provides a tunable, adaptive graph construction in which edges encode the non-local electronic overlap between atoms. This graph is then decomposed into a set of smaller, overlapping subgraphs, where the electronic structure of each of these subgraphs is solved for independently and in parallel using an external electronic-structure code.
SEDACS can be coupled to a variety of external electronic-structure solvers with minimal modifications to their software, enabling rapid adoption of the graph-based QMD approach. In this way, SEDACS can greatly extend the capability of existing electronic-structure packages by enabling stable QMD simulations of systems that were previously computationally inaccessible.
We demonstrate highly efficient and stable QMD simulations for chemically active systems with tens of thousands of atoms by interfacing SEDACS with an external Fortran-based electronic-structure code based on self-consistent-charge density functional tight-binding theory. 
\end{abstract}

\maketitle

\section{Introduction}

Molecular dynamics (MD) simulations play an important role in our understanding of chemical reactivity, structures,  materials properties, and biological functions \cite{RCar85, MAllen90, PCarloni02, JTse02, AVoter02, RIftimie05, marx_hutter_2009, MTuckerman10, MKarplus14}. Recent developments in machine‑learned (ML) interatomic potentials have enabled MD simulations to achieve near-first-principles accuracy while maintaining excellent computational efficiency and scalability \cite{WJia20, NFedik22, YWZhang25, MKulichenko24, YCOuyang25}. Nevertheless, the transferability of many ML interatomic potentials remains limited outside the chemical and configurational space covered by their training datasets. Moreover, because these models do not include the governing physical laws in an explicit manner, their predictions are often challenging to interpret, analyze, and generalize \cite{MKulichenko24, YCOuyang25, BKalita25}.
 
In contrast, quantum-mechanical molecular dynamics (QMD), where forces are derived from an underlying quantum-mechanical description of the electronic structure, provides physically transparent results and is broadly applicable. QMD can naturally capture phenomena such as bond breaking and interatomic charge transfer in chemically heterogeneous environments, including long-range Coulomb interactions and charge relaxations. However, its applicability is limited by the unfavorable (typically cubic) scaling of the electronic-structure calculations, which restricts accessible system sizes and simulation time scales.

First principles density functional theory (DFT) \cite{hohen,KohnSham65,RParr89,RMDreizler90} is often considered the gold standard. This approach offers a favorable balance between accuracy and computational cost, but its cubic scaling with system size typically bounds simulations to at most a few thousand atoms. Alternatively, the development of linear-scaling electronic-structure methods has enabled calculations of systems containing millions of atoms \cite{SGoedecker99, DBowler10, DBowler12, SMohr15, Fattebert2016MillionAtoms} but its broader applicability remains limited because accuracy is often reduced in ways that can be difficult to control systematically. In addition, the computational prefactor is often quite high, with linear-scaling benefits emerging only at very large system sizes beyond practical time or resource limits. These limitations are particularly challenging for QMD simulations.

Semiempirical approaches such as self-consistent-charge density-functional tight binding theory (SCC-DFTB) \cite{MElstner98,MFinnis98,TFrauenheim00,BHourahine20} significantly reduce computational cost and enable larger QMD simulations, albeit with some loss of accuracy compared to first principles DFT. Nevertheless, extending semiempirical QMD simulations to tens of thousands of atoms remains challenging, as traditional implementations still rely on cubically scaling solvers that become prohibitively expensive for large systems.

A recent advance addressing these limitations is graph-based linear-scaling electronic-structure theory \cite{ANiklasson16,CNegre_2022}. Graph-based linear-scaling electronic structure theory and its subsequent theoretical analysis, applications, and implementations \cite{Djidjev16,MLAss18,Djidjev19,MLass20, MKulichenko25} exploit the electronic locality in non-metallic systems by representing electronic connectivity as a sparse graph and reformulating the electronic structure calculations as collections of independent dense matrix operations on graph-partitioned subdomains. This enables natural parallelism with rigorously controllable accuracy, i.e.\ combining the strengths of divide-and-conquer \cite{WYang91,WYang95,Kitaura1999FMO, TOzaki06,VQuan19,YNishimoto21} and numerically thresholded sparse matrix methods \cite{SGoedecker99,APalser98,ANiklasson02,DBowler12,EHRubensson14,Truflandier16,UBORSTNIK14,VWeber15,PPinski15,AKruchinina16,JFinkelstein21}. However, applying linear-scaling approaches to QMD is challenging because approximations like numerical thresholding or radial cutoffs can hinder self-consistent-field (SCF) convergence and lead to non-conservative forces and instabilities. To overcome this, we integrate graph-based linear-scaling electronic-structure theory with time-reversible extended Lagrangian Born–Oppenheimer molecular dynamics (XL-BOMD) \cite{ANiklasson08,THirakawa17} using a shadow-potential formulation \cite{ANiklasson21b,ANiklasson23}. In this shadow MD framework, auxiliary electronic degrees of freedom are propagated as extended dynamical variables that remain close to the exact instantaneous ground-state charge density. This eliminates the need for iterative SCF optimization and maintains long-term energy stability by evolving the system on an approximate shadow potential energy surface that is virtually indistinguishable from the exact potential that would have been obtained via a costly iterative, brute-force procedure.

Despite these methodological advances, the practical adoption of graph-based QMD has been limited by the lack of flexible software infrastructure capable of interfacing graph-based electronic-structure algorithms with existing, production-level electronic-structure codes, particularly those written in compiled languages. Generalizing the approach to new methods, hardware architectures, or large-scale applications is therefore difficult.

This article introduces the Scalable Ecosystem, Driver, and Analyzer for Complex chemistry Simulations (SEDACS), a modular Python-based framework designed to bridge this gap. SEDACS provides a flexible environment for integrating graph-based linear-scaling electronic-structure theory, extended Lagrangian shadow MD, and scalable parallel execution interfaces with external electronic-structure codes. As a demonstration, we couple SEDACS with the Los Alamos Transferable Tight-binding for Energetics (LATTE) code, an SCC-DFTB-based electronic-structure solver written in Fortran \cite{LATTE, Bock2008-bp,Perriot2018-cg} in order to support graph-based techniques and shadow molecular dynamics through a lightweight Python interface, without requiring extensive modification to the underlying Fortran source code. In addition, we introduce a tunable dynamic graph adaptation strategy that updates graph connectivity at each integration step, improving error control and computational efficiency. Working together, these developments enable stable, scalable QMD simulations of chemically complex systems containing tens of thousands of atoms, substantially expanding the scope of established electronic structure codes.

\section{Graph-Based Linear-Scaling Electronic-Structure Theory}

In graph-based linear-scaling electronic-structure theory, the quantum-mechanical eigenvalue problem is reformulated as a sparse matrix Fermi-operator expansion of the density matrix, defined on an electronic connectivity graph, where sparsity is controlled by a numerical threshold \cite{ANiklasson16, CNegre_2022}. For non-metallic systems, the electronic overlap is naturally sparse with exponentially decaying Wannier functions, which enables linear scaling by decomposing the global problem into independent, dense computations on small, overlapping subgraphs. In SEDACS this approach provides natural parallelism, similar to a divide-and-conquer-like strategy, and allows systematically controllable accuracy through adjustable numerical thresholding as in a sparse matrix method.

\subsection{Matrix Functions on a Graph}
\label{functiom_on_graph}

Of key importance in graph-based linear-scaling electronic-structure theory is the concept of a function on a graph, denoted by $f_G(\mathbf{X})$ or $f(\mathbf{X})\big \vert_G$ \cite{ANiklasson16,CNegre_2022,MKulichenko25}. By definition, a function on a graph requires a step-by-step polynomial expansion of the matrix function $f(\mathbf{X})$, where each intermediate operation retains only matrix elements that correspond to the edges and vertices of the graph ${\bf G}$. The graph ${\bf G}$ can be determined from a global numerical threshold, $\tau$, where any element during the polynomial matrix expansion of $f(\mathbf{X})$ whose absolute value is above $\tau$ is kept throughout the whole expansion. These elements then form the edges of ${\bf G}$. 

The graph $\mathbf{G}$ is partitioned into smaller overlapping subgraphs, $\{{\mathbf g}^{(i)}\}$, consisting of non-overlapping core parts (c) and overlapping halos (h). These subgraphs can then be used to extract the corresponding principal submatrices, $\{\mathbf{x}^{(i)}\}$, of $\mathbf{X}$ for each subgraph $i$ \cite{ANiklasson16}.
A key observation behind graph-based linear-scaling electronic-structure theory is the equivalence between a sparse matrix function on a graph, $f_\mathbf{G}({\bf X})$ (i.e.\ the globally thresholded step-by-step matrix function expansion) and a collection of the core contributions of the dense matrix functions, $f_c(\mathbf{x}^{(i)})$, computed over the principal submatrices of $\mathbf{X}$, i.e., \
\begin{align}
f_G(\textbf{X}) &  = \Bigl\{f_c(\textbf{x}^{(i)})\Bigr\}_{\text{collect}}.
\end{align}
This relation defines a one-to-one mapping between a numerically thresholded sparse matrix algebra and a divide-and-conquer-like approach, offering a natural way to parallelism and linear scaling complexity, while maintaining control of the error with an adjustable numerical threshold. SEDACS uses this one-to-one relation to provide scalable electronic structure calculations with tunable accuracy.

More details of matrix functions on a graph and the equivalent calculation over partitioned subgraphs are presented in \textit{SI Appendix}, Section A and in Refs.\ \cite{ANiklasson16, CNegre_2022, MKulichenko25}.

\subsection{Density Matrix Evaluation on Partitioned Graphs}
\label{dm_on_graph}

In graph-based linear-scaling electronic-structure theory, the matrix representation of an operator (e.g., the Hamiltonian or the single-particle density matrix) can be viewed as a graph, where vertices represent atoms or orbitals, and edges denote the corresponding electronic connectivity or overlap. SEDACS uses a coarse-grained atom-based connectivity representation (rather than an orbital-based one), which enhances computational efficiency and substantially reduces the memory footprint and inter-node communication needed to construct and maintain the graph representation. For simplicity, in the discussion below we will assume an orthonormal basis-set representation, where the basis-set overlap matrix, ${\bf S}$, is equal to the identity matrix ${\bf I}$, ignoring the change of basis transformation between an atomic-orbital basis and the orthonormalized representation (see Ref.\ \cite{CNegre_2022} for details).

The matrix function on a graph that we use in graph-based linear-scaling electronic-structure theory is the Fermi function applied to the effective (Kohn-Sham) Hamiltonian, $\mathbf{H}$, which yields the single-particle density matrix, $\mathbf{D}$. This matrix, $\mathbf{D}$, then determines the electronic structure and associated observables, such as the charge density, total electronic energy, dipole moments, and atomic forces. Assuming an orthonormalized basis-set representation, a density matrix calculated on a graph, $\textbf{D}_{G}$, can be expressed in terms of the Fermi-operator expansion,
\begin{align}
    \textbf{D}_G  = f_G(\mathbf{H})  &\equiv \Bigl[ e^{\beta(\textbf{H}-\mu\textbf{I})} + \textbf{I}\Bigr]^{-1} \bigg|_{G} \approx \sum_{k} a_{k}T_{k}(\textbf{H}) \bigg|_G.
\end{align}
for some set of expansion coefficients $\{a_k\}$ and basis functions $\{T_k(\bf H)\}$. For example, this could be a Chebyshev expansion, in which $T_k(\bf H)$ is the $k$-th Chebyshev polynomial. Here $\beta = 1/(k_BT)$ denotes the inverse electronic temperature and $\mu$ is the chemical potential, determined such that the trace of the density matrix equals the number of occupied states, i.e.\ ${\rm Tr}[\mathbf{D}] = N_{\rm occ}$.

We can now recast a density matrix calculated on a graph, $\mathbf{D}_G$, as a collection of core contributions from independent dense Fermi-operator expansions over the principal submatrices of the Hamiltonian, $\{\mathbf{h}^{(i)}\}$, determined by the core-halo subgraphs, $\{\mathbf{g}_i\}$, of $\mathbf{G}$. In this case,
\begin{align}
\mathbf{D}_G(\textbf{H}) & = \sum_{k} a_{k}T_{k}(\textbf{H}) \Big|_G 
= \Bigl\{f_c(\textbf{h}^{(i)})\Bigr\}_{\text{collect}}.
\end{align}

\begin{figure*}[htbp]
    \centering
    \includegraphics[width=0.99\textwidth]{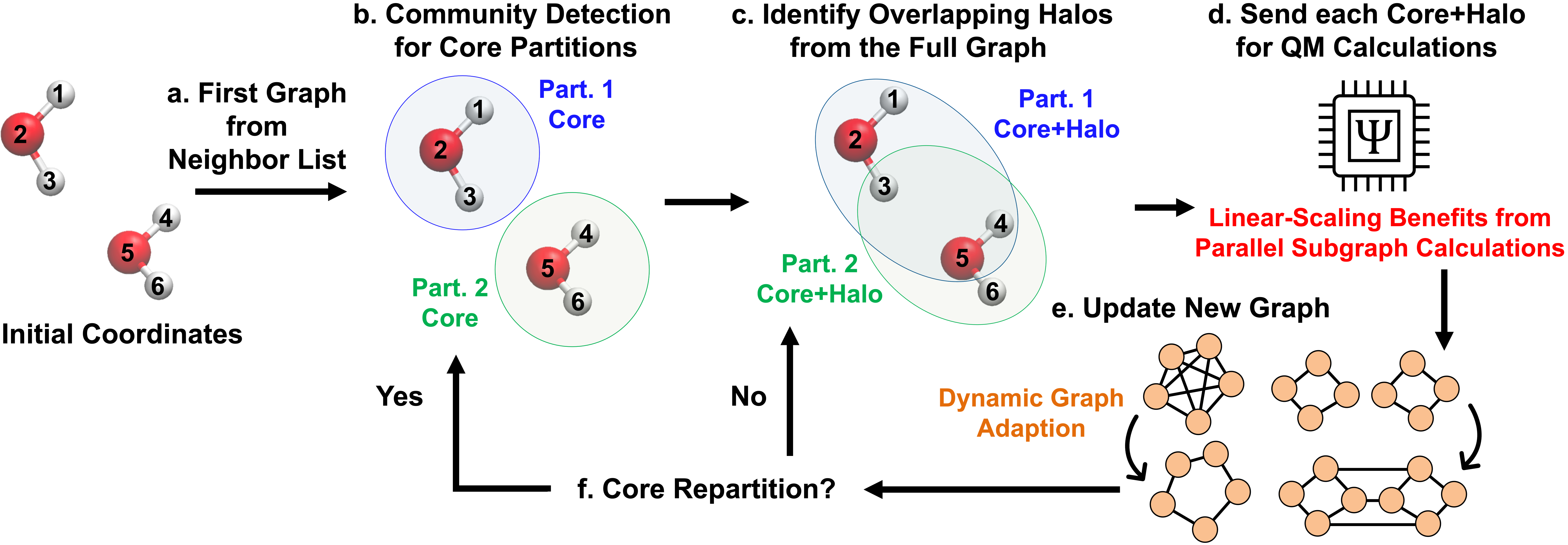}
    \caption{Schematic workflow of the adaptive graph partitioning in SEDACS. a.~Starting from the initial atomic coordinates of the full system (two water molecules), the graph is constructed from a  neighbor list. b.~The graph is then partitioned to define the non-overlapping core partitions (atoms 1-3 form the core partition 1 (part. 1), and atoms 4-6 form the core partition 2 (part. 2)), and c.~the associated halo regions are determined by including the neighboring atoms of each core part (atom 4-5 become the halo of partition 1 and atom 3 becomes the halo of partition 2). d.~Once the core–halo regions are established, quantum mechanical calculations for each partition can be performed in parallel, enabling linear-scaling efficiency for systems with sparse graph connectivity. e.~The resulting density matrices from all partitions are then assembled, including additional (possibly redundant) halo edges from a weighted distance graph (See Sec.\ \ref{dynamic_graph_adaption}), to construct a thresholded updated graph for the next step of the calculation. f.~The core repartitioning can be updated if necessary, e.g.\ as the atomic configuration changes significantly.}
    \label{graph_partition}
\end{figure*}

\noindent SEDACS exploits this equivalence to enable electronic-structure calculations that combine inherent parallelism and systematically tunable accuracy. The dense matrix algebra operations on each subgraph are independent and well-suited to execution on hardware accelerators such as GPUs. Moreover, the dense matrix expansions on the principal submatrices are agnostic to the particular scheme employed to evaluate the matrix functions, because it does not depend on any matrix sparsity, graph topology or numerical thresholds. The subgraph Fermi-operator expansions can therefore be performed using recursive Fermi-operator expansion schemes \cite{APalser98, ANiklasson02, EHRubensson14, JFinkelstein21, JFinkelstein22} or, alternatively, in their molecular orbital representations using direct diagonalization.

Apart from calculating the electronic structure, forces, energies, and related properties via graph-based, distributed density matrix calculations, SEDACS can also perform response calculations based on quantum perturbation theory on a partitioned graph. This enables computing the density matrix response to a perturbation in the Hamiltonian and is described in Ref.\ \cite{CNegre_2022}. Graph-based quantum perturbation theory is important for the kernel calculation in the shadow molecular dynamics formulation and for the low-rank kernel approximation and accelerated SCF optimization discussed in Sections \ref{ShadowMD} and \textit{SI Appendix}, Section B and C, respectively.

\subsection{Determining the Subgraphs}

In general, the data dependency graph for the Fermi-operator expansion, $\textbf{G}_{\tau}$, which can be tuned by a global numerical threshold, $\tau$, is not known \textit{a priori}. This presents a challenge, as knowledge of the graph is required to perform the very graph partitioning needed to enable the partitioned parallel computation.  Luckily, in iterative SCF calculations or in MD simulations, the data dependency graph can always be estimated from previous steps, including some additional (and possibly redundant) halo edges to enable an adaptive and tunable estimate of the data dependency graph.

Fig.\ \ref{graph_partition} illustrates the workflow underlying the adaptive graph-based electronic-structure framework implemented in SEDACS, which is used in the first SCF optimization in the initial MD step (a to f) and in the subsequent MD steps (b to f). The initial SCF procedure begins by estimating the first full graph from scratch using a distance-dependent neighbor list (Fig.\ \ref{graph_partition}a). This graph is then used to define a core partitioning over non-overlapping regions (Fig.\ \ref{graph_partition}b and c) using off-the-shelf community detection algorithms \cite{metis}, keeping overlapping halo regions small at little extra cost. Before SCF optimization, the initial partitioning is solely derived from neighbor lists, without incorporating any electronic information. Consequently, outer partitions of the halo may, in principle, cut through chemical bonds in extended or covalently connected systems (e.g., polymers). During the iterative SCF optimization, however, halo regions are adaptively expanded or contracted so that chemically relevant interactions are progressively recovered using adaptive, thresholded, weighted graphs. These graphs are defined not only by a distance-weighted neighbor list, but also by the atom-condensed density matrix, which captures the non-local electronic overlap. As a result, the graphs and their halo regions are dynamically adapted to the electronic connectivity (Fig.\ \ref{graph_partition}e) in contrast to, for example, a fixed radial cutoff divide-and-conquer approach.

A ground-state SCF optimization is only needed for the first time step for our graph-based shadow QMD simulations and it can be accelerated with a particular kernel-based approach adapted to the distributed graph partitioning as discussed in \textit{SI Appendix}, Section C. In each subsequent MD step (or in the initial SCF optimization), the approximate graph is used to partition the density matrix calculation into independent subgraphs that are evaluated in parallel (Fig.\ \ref{graph_partition}d). In the following MD step (or SCF iteration), new data dependency graphs are then estimated from a tunable dynamic graph adaption scheme in Sec.\ \ref{dynamic_graph_adaption}. This step can form new additional halo edges, while connections below a chosen threshold, $\tau$, are removed (Fig.\ \ref{graph_partition}e). In the initial SCF iterations, edges are typically only added and not removed, whereas edges are both added and removed automatically as the atoms move and the electronic overlap changes during a QMD simulation. The core partitioning can be reused between time steps in order to avoid unnecessary repartitioning overhead. However, if structural or electronic changes significantly increase the halo sizes, a new core partitioning may be introduced to restore efficiency (Fig.~\ref{graph_partition}f). Such repartitioning events are infrequent and their cost amortized over the many time steps of an MD simulation.

For metallic systems, the adaptive scheme may ultimately produce an all-to-all connectivity graph due to the delocalized all-to-all nature of itinerant electrons. In this case, each core and halo effectively spans the entire system, eliminating both natural parallelism and linear scaling. Nevertheless, the method remains accurate and yields the correct electronic structure. This contrasts with divide-and-conquer approaches that impose a fixed radial cutoff, and while they retain favorable scaling and parallel efficiency, they can produce qualitatively incorrect results. 
Although electronic states may be delocalized over extended regions, the presence of a gap ensures exponential decay of the density matrix, allowing the adaptive graph to remain sparse and the method to retain substantial parallel efficiency. Owing to its dynamic graph adaptation, SEDACS can therefore provide reliable QMD results even when substantial changes in electronic structure occur, including chemical reactions or, in an extreme case, insulator-to-metal transitions.

\section{Shadow Molecular Dynamics}

One of the main challenges in QMD simulations is obtaining the SCF solution of the relaxed electronic ground state, which determines the Born-Oppenheimer potential energy and interatomic forces in each time step. This typically involves an iterative procedure of repeated solutions of non-linear quantum-mechanical eigenvalue problems that need to be solved with tight convergence to ensure accurate, conservative forces and to prevent unphysical behavior, broken time-reversal symmetry, instabilities and energy drift \cite{DRemler90,PPulay04,ANiklasson06}. The iterative SCF optimization is the most computationally expensive step, often limiting the accessible time and length scales of the simulations. Our graph-based SEDACS framework transforms the quantum-mechanical eigenvalue problems into a partitioned electronic-structure calculation, which can be solved in parallel with linear-scaling complexity. However, because of the numerical thresholding used for the estimated data dependence, the graph-based approach is approximate. These approximations often make SCF convergence challenging (or even impossible) and errors in the forces may lead to instabilities and unphysical trajectories that potentially invalidate the QMD simulations. To avoid these shortcomings, SEDACS combines graph-based electronic-structure theory with a shadow MD formulation within the framework of extended-Lagrangian Born--Oppenheimer molecular dynamics (XL-BOMD) \cite{ANiklasson17,ANiklasson20,ANiklasson21b,ANiklasson23}. This shadow MD approach removes the requirement of an iterative SCF optimization. In this way, we can both reduce the computational cost and increase the stability. The key ideas underlying the shadow MD used in SEDACS are twofold: 1) to propagate electronic degrees of freedom as dynamical variables in the spirit of Car-Parrinello molecular dynamics \cite{RCar85}, while ensuring that the electrons stay close to their instantaneous ground state; and 2) to run the dynamics on an approximate shadow potential energy surface obtained through direct and exact equilibration of an approximate (linearized) energy functional. This approach avoids both the potential instabilities and the costs associated with the iterative SCF solutions used to converge an ``exact'' conventional Born-Oppenheimer potential energy surface. With the latest shadow potential formulation of XL-BOMD,
we are even able to simulate challenging, charge-sensitive, and complex, chemically active systems without performing any SCF iterations.

The shadow XL-BOMD formulation is a key ingredient in SEDACS that makes our graph-based approach to stable, large-scale QMD simulations possible.

\subsection{Shadow Potential Energy}

In DFT, the ground-state electron density, $\rho_{\mathrm{min}}(\mathbf{r})$, and the corresponding Born-Oppenheimer potential energy, $U_{\rm BO}(\mathbf{R})$, are given by
\begin{equation}
    \rho_{\mathrm{min}}(\mathbf{r}) = \arg\min_{\rho} \biggl\{ E(\mathbf{R}, \rho)\;\bigg|\; \int \rho(\mathbf{r})\, d\mathbf{r} = N_e \biggr\},
    \label{exact_min}
\end{equation}
\begin{equation}
    U_{\rm BO}(\mathbf{R}) = E(\mathbf{R}, \rho_{\mathrm{min}}).\label{BO_Pot}
\end{equation}
Here, \( E(\mathbf{R}, \rho) \) denotes the {\em non-linear} density functional of the electronic energy for a given set of nuclear positions, \( \mathbf{R} \), and electron density, \( \rho \), including the ion-ion repulsion terms. The Born-Oppenheimer potential, \( U_{\rm BO}(\mathbf{R}) \), is obtained from the lowest stationary ground-state solution, \( \rho_{\mathrm{min}} \), determined by minimizing the energy over all physically relevant electron densities that integrate to the prescribed number of electrons, \( N_e \). In Kohn-Sham DFT we use a molecular orbital ansatz, $\{\psi_i({\bf r})\}$, for the electron density, where $\rho(\mathbf{r}) = \sum_i f_i |\psi_i({\bf r})|^2$ together with the Kohn-Sham expression for $E({\bf R},\rho)$ \cite{KohnSham65}. In this case the constrained equilibration in Eq.\ (\ref{exact_min}) leads to a {\em non-linear} quantum-mechanical eigenvalue problem that needs to be solved using an iterative SCF solver in each time step. In Kohn-Sham DFT we can also account for fractional occupation numbers, $f_i \in [0,1]$, by including an additional entropy term to account for finite electronic temperature effects \cite{RParr89}. In this case our Born-Oppenheimer-like approximation corresponds to an instantaneously thermally equilibrated electronic structure and the Born-Oppenheimer potential becomes a free energy surface.
The SCF optimization is computationally demanding and insufficient convergence may lead to unphysical trajectories and  instabilities in QMD simulations. Shadow MD is designed to avoid these problems. It is based on the backward error analysis of shadow Hamiltonians in geometric integration, enabling stable trajectories without the need for tightly converged iterative solvers.

To construct a shadow potential that avoids the nonlinear constrained minimization problem in Eq.~(\ref{exact_min}), we first introduce an approximate shadow energy functional, \( \mathcal{E}(\mathbf{R}, \rho, n) \), obtained by linearizing the exact energy functional \( E(\mathbf{R}, \rho) \) with respect to the electron density around a trial ground-state density, \( n(\mathbf{r}) \), where
\begin{equation}
\begin{aligned}
    \mathcal{E}(\mathbf{R}, \rho, n) &= E(\mathbf{R}, n) \\
    &+ \int \big(\rho({\bf r}) - n({\bf r}) \big)\frac{\delta E({\bf R},\rho)}{\delta \rho({\bf r})}\Bigg \vert_n d{\bf r}.
\end{aligned}
\label{Shadow_Energy}
\end{equation}
By performing a constrained equilibration of this approximate shadow energy functional, we then obtain the fully relaxed \( n \)-dependent ground-state density, \( \rho_{0}[n](\mathbf{r}) \), and the corresponding \(n\)-dependent shadow Born-Oppenheimer potential, \( \mathcal{U}_{\rm BO}(\mathbf{R}, n) \), i.e.\
\begin{equation}
    \rho_{0}[n](\mathbf{r}) = \arg\min_{\rho} \biggl\{ \mathcal{E}(\mathbf{R}, \rho, n)\;\bigg|\; \int \rho(\mathbf{r})\, d\mathbf{r} = N_e \biggr\},
    \label{shadow_min}
\end{equation}
\begin{equation}
    \mathcal{U}_{\rm BO}(\mathbf{R}, n) = \mathcal{E}(\mathbf{R}, \rho_{0}[n], n).
\end{equation}
With the minimization or ground-state equilibration we mean the lowest stationary solution. Due to linearization, the constrained equilibration in Eq.\ (\ref{shadow_min}) leads to a {\em linear} quantum-mechanical eigenvalue problem that can be solved directly in a single step. No iterative SCF solver is needed.
If \( \mathbf{H}[n] \) is the $n$-dependent Kohn-Sham Hamiltonian matrix (determined by the shadow energy functional in Eq.\ (\ref{Shadow_Energy}) from $\delta {\cal E}\big/ \delta n$) using some orthonormalized basis set representation, $\{\phi_i({\bf r})\}$, then this single-step solution can be calculated through the density matrix, ${\bf D}[n]$, where
\begin{equation}
    \mathbf{D}\big(\mathbf{H}[n]\big) = \Bigl[ e^{\beta(\mathbf{H}[n] - \mu \mathbf{I})} + \mathbf{I} \Bigr]^{-1},
\end{equation}
with the electron density,
\(\rho[n]({\bf r}) = \sum_{ij}D_{ij}[n]\phi_i({\bf r})\phi_j({\bf r})\).
The chemical potential, $\mu$, is chosen to ensure the correct occupation at the inverse electronic temperature, $\beta$. For simplicity, we neglect accounting for a non-orthonormalized basis set and assume that the overlap ${\bf S} = {\bf I}$. 

This linearized Fermi-operator formulation is used in SEDACS and naturally connects to both the shadow density functional approach and the graph-based linear-scaling electronic-structure methods discussed in Section~\ref{dm_on_graph}.

\subsection{Extended Lagrangian Born-Oppenheimer Molecular Dynamics}
\label{ShadowMD}

The shadow Born-Oppenheimer potential, \( \mathcal{U}_{\rm BO}(\mathbf{R}, n) \), is obtained from the constrained minimization of the linearized shadow energy functional, \( \mathcal{E}(\mathbf{R}, \rho, n) \). The difference to the ``exact'' Born-Oppenheimer potential, $U_{\rm BO}(\textbf{R})$, scales as

\begin{equation}
    \vert {\cal U}_{\rm BO}(\textbf{R}, n) - U_{\rm BO}(\textbf{R})\vert \propto |\rho_{0}[n]-n|^2, 
    \label{UErr}
\end{equation}
where the residual, $\rho_{0}[n]-n$, disappears at the exact ground state when $n = \rho_{\rm min} = \rho_0[n]$.
To ensure that the density, \( n(\mathbf{r}) \), remains close to the instantaneous ground-state density as the nuclear positions evolve during MD simulations, we propagate \( n(\mathbf{r}) \) as an extended dynamical field variable driven by a harmonic oscillator with a harmonic potential following the ground-state density, $\rho_0[n]$. This additional dynamics can be formulated within an extended Lagrangian framework \cite{ANiklasson08,ANiklasson21b}. The dynamics is then given from the Euler-Lagrange equations of motion. In an adiabatic Born-Oppenheimer-like limit, where we assume that the extended dynamical charge density is fast compared to the nuclear motion, we get the coupled equations of motion, 
\begin{align}
    M_I \ddot{\mathbf{R}}_I &= \left. -\nabla_{\mathbf{R}_I} \mathcal{U}_{\rm BO}(\mathbf{R}, n) \right|_{n}, \label{NEQ} \\
    \ddot{n}(\mathbf{r}) &= - \omega^2 \int K(\mathbf{r}, \mathbf{r}') 
    \bigl( \rho_0[n](\mathbf{r}') - n(\mathbf{r}') \bigr)\, d\mathbf{r}', \label{HW}
\end{align}
where \( \{M_I\} \) denote the atomic masses.
The first equation, Eq.\ (\ref{NEQ}), corresponds to Newton's equations of motion for the nuclear degrees of freedom, with forces obtained from the gradient of the approximate shadow potential, while holding the dynamical variable density, \( n(\mathbf{r}) \), fixed. The second equation, Eq.\ (\ref{HW}), describes a linear dynamics for the electron density, \( n(\mathbf{r}) \), which evolves in a harmonic well centered at \( \rho_0[n](\mathbf{r}) \).
Here the kernel, \( K(\mathbf{r}, \mathbf{r}') \), is defined as the inverse Jacobian, $K = J^{-1}$, where
\begin{equation}
    J(\mathbf{r}, \mathbf{r}') = 
    \frac{\delta \bigl( \rho_0[n](\mathbf{r}) - n(\mathbf{r}) \bigr)}
    {\delta n(\mathbf{r}')}.
\end{equation}
The kernel keeps $n({\bf r})$ close to the exact ground-state density, $\rho_{\rm min}$, which further stabilizes the dynamics. As a result, the density, \( n(\mathbf{r}) \), closely tracks the true ground state and the error in the shadow potential (Eq.~(\ref{UErr})) remains small.  This is particularly important for charge-sensitive solutions involving chemical reactions.

The shadow MD above is described in terms of the electron density, \(n({\bf r} \)), but other representations of the extended electronic degrees of freedom can alternatively be used, for example, the density matrix or the wavefunctions \cite{ANiklasson20b,MKulichenko25,PSteneteg10}.

In SEDACS, the equations of motion given by Eqs.\ (\ref{NEQ}) and (\ref{HW}) are integrated using a modified velocity-Verlet scheme \cite{ANiklasson09,PSteneteg10,GZheng11}. The integration scheme includes a weak dissipation that prevents the accumulation of numerical noise due to the adiabatic approximation, and keeps the dynamical charge density synchronized with the nuclear motion. The electronic degrees of freedom are initialized by setting \( n(\mathbf{r}) = \rho_{\mathrm{min}}(\mathbf{r}) \), which requires a full SCF optimization of the ground-state density, as in Eq.\ (\ref{exact_min}), but only at the initial time step. Nevertheless, even a single SCF optimization of a large complex system that is chemically unstable, including tens-of-thousands of atoms, can be a major challenge. In SEDACS we have developed a kernel-based quasi-Newton scheme that provides a robust and competitive alternative to SCF-acceleration schemes such as Pulay mixing \cite{PPulay80,PPulay82}. This approach is discussed and demonstrated in the \textit{SI Appendix}, Section C.

The accuracy of the shadow potential depends on the size of the integration time step, \( \delta t \), to the fourth order \cite{ANiklasson17}, such that
\begin{align}
    \vert {\cal U}_{\rm BO}(\textbf{R}, n) - U_{\rm BO}(\textbf{R})\vert
    = \mathcal{O}\!\left( \bigl| \rho_0[n] - n \bigr|^2 \right)
    \propto \delta t^4 .
\end{align}
In practice, this sampling error is negligible compared to the local truncation error associated with the Verlet integration scheme. In general, SEDACS uses the same integration time step as in conventional direct Born-Oppenheimer MD simulations \cite{ANiklasson21b}. However, for chemically active systems, the time step typically has to be reduced to capture the more rapid changes in the electronic structure.

The kernel in Eq.~(\ref{HW}) can often be approximated by a scaled delta function,
\(
K(\mathbf{r}, \mathbf{r}') = -c\, \delta(\mathbf{r} - \mathbf{r}'), \; c \in [0,1].
\)
However, for complex chemically active systems with sensitive or unstable charge distributions, this approximation can be insufficient. In this case, SEDACS uses a preconditioned low-rank Krylov subspace approximation of the kernel, based on graph-based quantum-response calculations \cite{ANiklasson20, CNegre_2022}. This mitigates the computational cost associated with evaluating the full kernel \cite{CNegre_2022}, as detailed in the \textit{SI Appendix}, Section B.

\section{SEDACS Software Design}

SEDACS is designed as a modular and extensible framework implemented in Python to enable massively parallel atomistic electronic-structure simulations, allowing seamless integration with a diverse range of available and emerging external electronic-structure packages at different levels of theory.

\begin{figure}[htbp]
    \centering
    \includegraphics[width=0.33\textwidth]{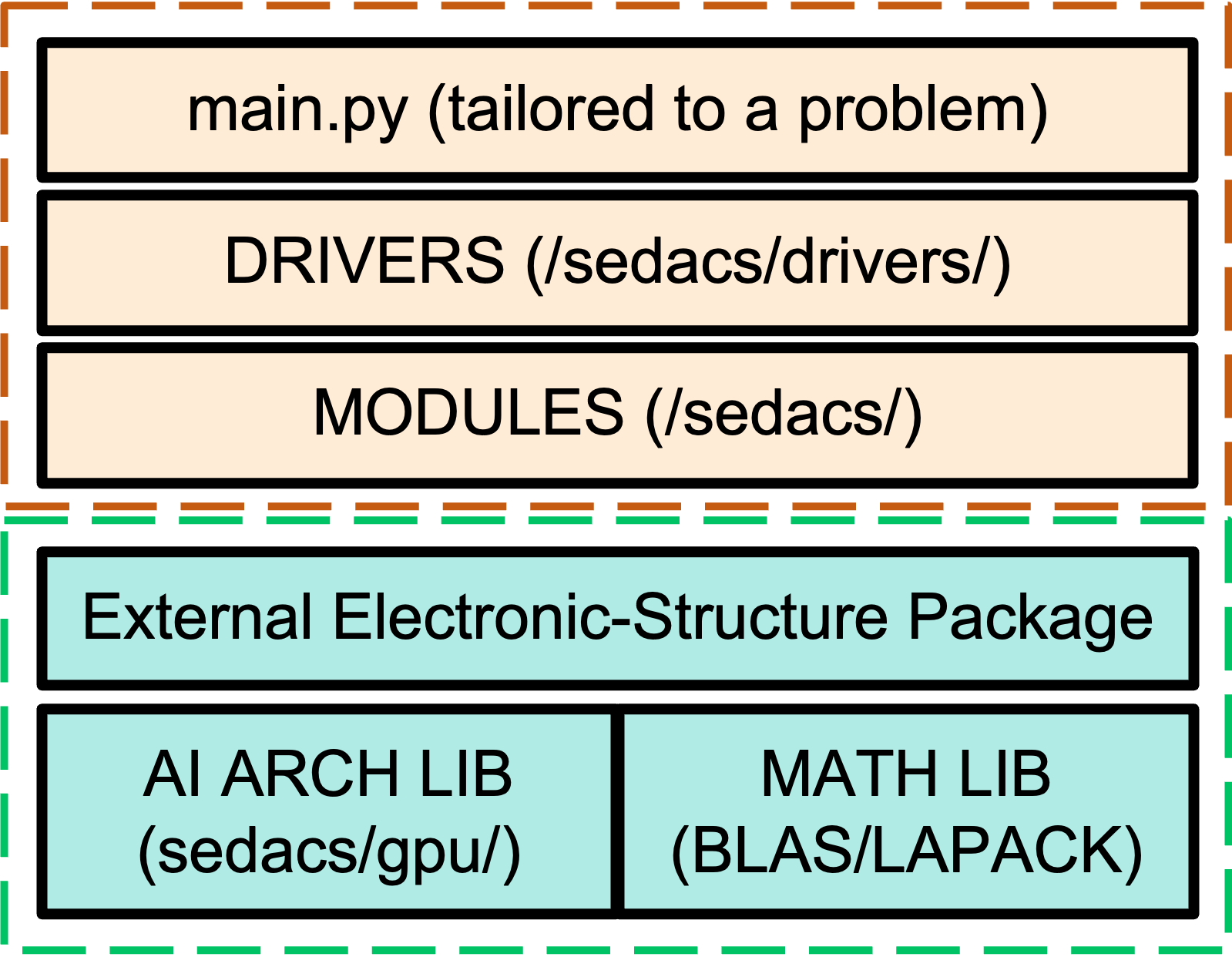}
    \caption{Software stack illustrating the primary application, the SEDACS drivers and modules, the external electronic-structure packages, and any library dependencies.}
    \label{sedacs_stack}
\end{figure}

As shown in Fig.~\ref{sedacs_stack}, SEDACS is organized as a stack of layered components: main files customized to a particular problem, drivers responsible for control and orchestration of modules and functions; external electronic-structure packages serving as computational engines for underlying electronic-structure theory implementations; and optional accelerator libraries for boosting the performance of external electronic-structure packages.

\subsection{SEDACS Overview}

\begin{figure*}[t]
    \centering
    \includegraphics[width=0.99\textwidth]{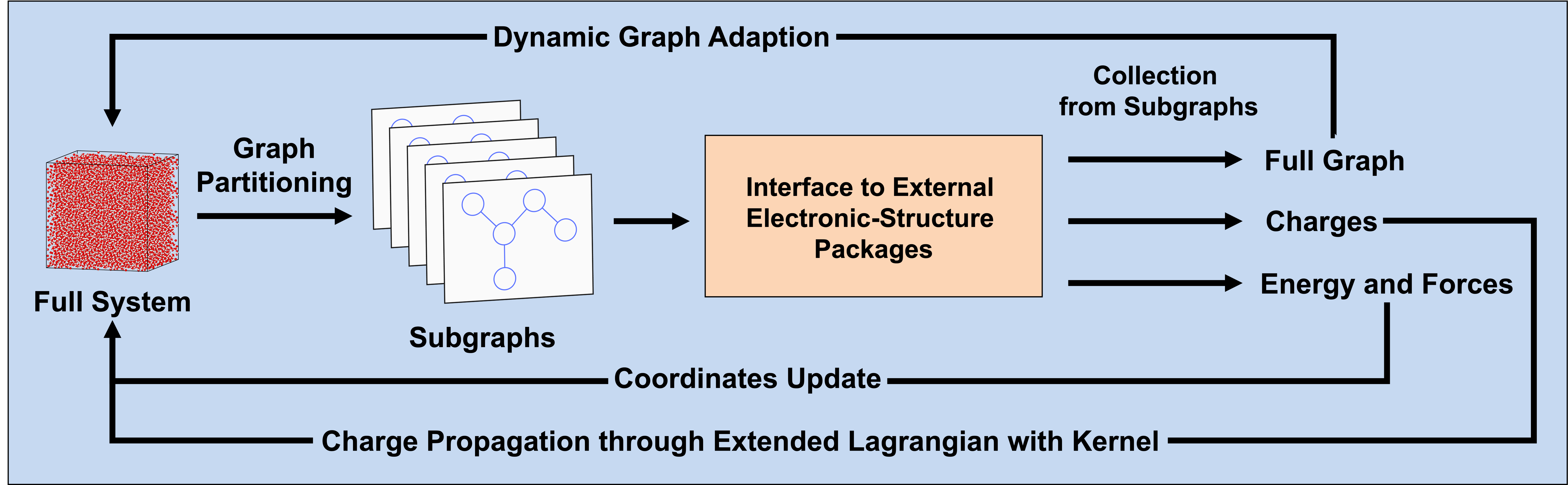}
    \caption{Overview of the SEDACS framework. User-defined inputs, including the full atomic system and simulation parameters, are first passed to SEDACS. Based on this information, SEDACS constructs a graph representation of the system and partitions it into subgraphs. The resulting subsystems are then dispatched in parallel to external electronic-structure packages for electronic-structure calculations. Upon completion, electronic quantities—including the updated graph connectivity, charges, total energy, and atomic forces—are assembled to recover the full-system description. Within SEDACS, modular components subsequently perform dynamic graph adaptation to generate an updated graph for the next iteration, while atomic coordinates and electronic degrees of freedom are propagated within the XL-BOMD framework.}
    \label{sedacs_overview}
\end{figure*}

SEDACS enables scaling by separating high-level simulation control from lower-level electronic-structure solvers. The top level manages the full physical system, simulation control parameters, and analysis tasks to tailor simulation workflows. This layer communicates with SEDACS drivers, which coordinate the graph construction and partitioning, electronic-structure evaluations, updating the positions and velocities for MD simulations, and data collection. Fig.~\ref{sedacs_overview} illustrates the control and data flow in the SEDACS framework, including the points where SEDACS connects to a chosen external electronic-structure code.

The core of SEDACS is a collection of reusable modules that implement graph-based linear-scaling electronic-structure theory, extended Lagrangian Born–Oppenheimer molecular dynamics, updates to the response kernel, and dynamic graph adaptation. These modules are agnostic to the specific electronic-structure solver and operate on well-defined data structures, such as graphs and density matrices. The modular design allows SEDACS to interface smoothly with multiple back-end external electronic-structure codes while maintaining a consistent high-level workflow. Optional accelerator libraries can be incorporated at this layer to enhance performance with hardware accelerators, without modifying the core simulation logic. Included with SEDACS is a custom-designed accelerator library that uses GPUs and AI-hardware (e.g. Nvidia Tensor cores) for costly dense computations such as Fermi-operator expansions and construction of the inverse overlap matrix \cite{AHabib24,JFinkelstein21}. 

Connecting external electronic-structure codes to SEDACS requires minimal modification of the external “engine” codes, as they are treated as computational back ends. In the present work, we demonstrate SEDACS using an external Fortran electronic-structure code, LATTE \cite{LATTE}, based on SCC-DFTB theory \cite{MElstner98,MFinnis98,TFrauenheim00,BHourahine20}. LATTE is compiled as a shared library and accessed through a lightweight Python interface using the ctypes module. Memory for key data structures, such as atomic coordinates, Hamiltonians, and density matrices, are allocated on the Python side as NumPy arrays, and raw pointers are passed to LATTE for computation. This approach minimizes data duplication while preserving compatibility with external codes, though at the cost of potential inefficiencies associated with Python memory management.

Overall, the architecture shown in Fig.~\ref{sedacs_overview} highlights how SEDACS decouples algorithmic innovation from the electronic-structure code development. By allowing both graph-based linear-scaling electronic structure theory and shadow molecular dynamics to be layered on top of established electronic-structure packages, our design significantly lowers the barrier to adopting scalable graph-based QMD techniques and facilitates rapid experimentation with new methods, solvers, and hardware platforms.

\subsection{Interface between SEDACS and External Electronic-Structure Package }

\begin{figure}[htbp]
    \centering
    \includegraphics[width=0.49\textwidth]{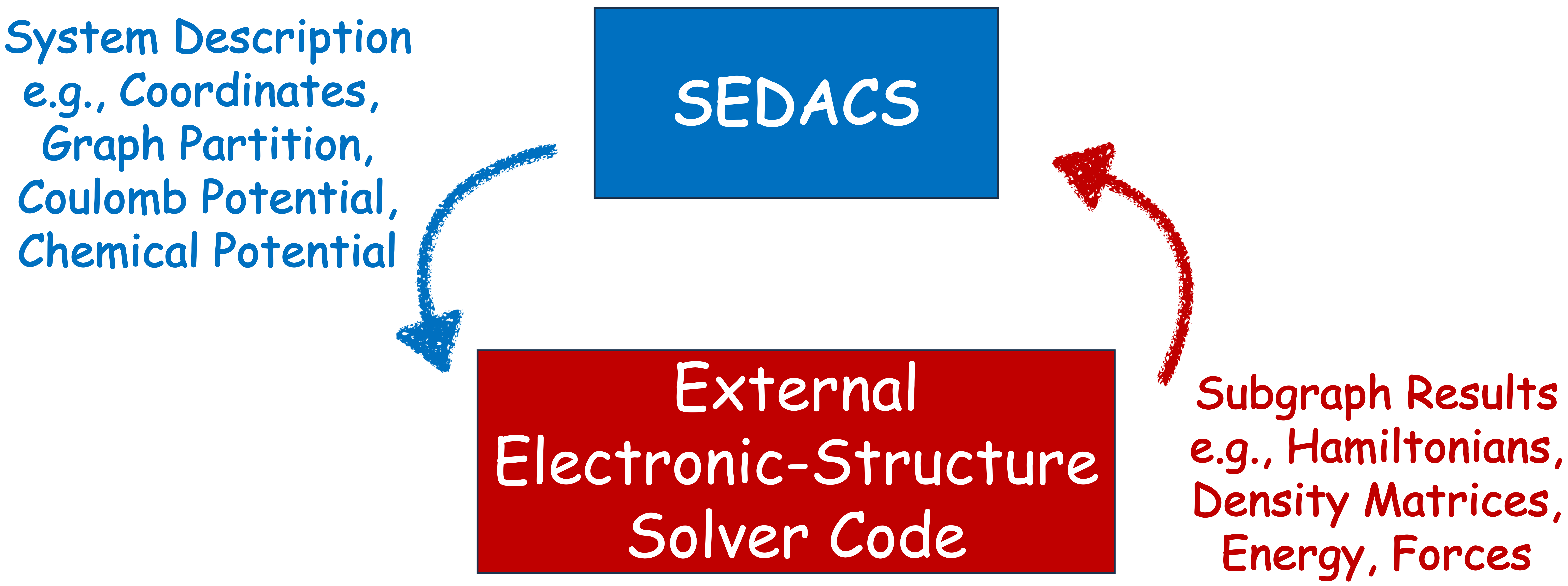}
    \caption{The interface between SEDACS and an external electronic-structure code and the flow of data between the two components. SEDACS manages the top level description of the full system, performs the necessary operations on the data, and decomposes it into subsystems that are dispatched in parallel to external electronic-structure packages. Once electronic-structure package complete these calculations, the resulting electronic-structure operators and observables are returned to SEDACS where they are collected, organized, and assembled to reconstruct the full-system representation.}
    \label{sedacs_qc_interface}
\end{figure}

To highlight the clear separation of responsibilities between high-level workflow management and low-level electronic structure evaluation, Fig.~\ref{sedacs_qc_interface} shows the interface between SEDACS and external electronic-structure packages. In this design, SEDACS serves as the central driver that manages simulation state, graph partitions, and global variables, while external electronic-structure packages are treated as a computational back-end responsible for evaluating electronic structure operators and observables. This separation allows SEDACS to perform large-scale simulations without requiring invasive and time-consuming modifications to the underlying electronic-structure code.

At each MD time step, SEDACS provides external electronic-structure packages with the essential input data required for electronic structure calculations. These inputs include atomic coordinates, atom types, lattice vectors, orbital indices, graph partitions defining core and halo regions, and global Coulomb and chemical potentials. The graph-partition information is particularly important because it specifies the subset of atoms and orbitals associated with each graph-based subproblem, enabling external electronic-structure package to operate on smaller, dense matrix blocks rather than the full system. All input data are prepared and managed on the SEDACS side, ensuring consistency across partitions and time steps.

Electronic-structure packages accept the communicated input data and returns a comprehensive set of electronic-structure outputs to SEDACS, including,  eigenvalues and eigenvectors, density matrices, atomic charges, total energies, and atomic forces for the different subgraphs. These quantities are used by SEDACS to update the graph topology, propagate the extended electronic degrees of freedom within the XL-BOMD framework, and advance the nuclear equations of motion. By returning density matrices rather than only forces or energies, the interface enables dynamic graph adaptation and kernel updates that are essential for maintaining accuracy and stability in graph-based QMD simulations.

\section{Simulations} 

This work presents the SEDACS framework and here we demonstrate its capabilities through simulations highlighting key components, including a challenging large-scale QMD simulation of a complex chemically active system. All simulations used the external electronic-structure code LATTE, based on SCC-DFTB theory \cite{LATTE}.

\subsection{Dynamic Graph Adaption} 
\label{dynamic_graph_adaption}

As atomic configurations evolve during QMD simulations, the graph must be able to expand or contract dynamically in response to changes in the electronic connectivity between the atoms. The exact electronic connectivity or graph is never known {\em a priori}. We therefore always include additional and possibly redundant edges to maintain accuracy and to allow new electronic overlap to form in the halo as the system evolves \cite{ANiklasson16,CNegre_2022}. However, to balance efficiency and physical fidelity, we need to use the smallest feasible halo extension for each subgraph.

SEDACS estimates the contractions and expansions of electronic connectivity from the numerically thresholded weighted graph, ${\bf G}_\tau$, formed by a weighted distance graph ${\bf G}^N$ whose edge weights given by
\begin{equation}
{\bf G}^N_{ij} = \exp(-\alpha R^{2}_{ij})\;,
\label{neigh_mat}
\end{equation}
are determined by the exponential decay of interatomic distances, $R_{ij}$, and an atom-condensed density matrix graph, ${\bf G}^D_{ij}$, given from the previous time step. The atom-condensed density matrix graph is derived from selecting the maximum element over each atom-atom block of the orbital-based density matrix. The connectivity graph, ${\bf G}_\tau$, is then determined by the numerically thresholded combined graph, 
\begin{equation}
    {\bf G}_\tau = \big( {\bf G}^N{\bf G}^D + {\bf G}^D{\bf G}^N\big) \big \vert_\tau. \label{Graph_Adaptation}
\end{equation}
This combined graph introduces new edges by linking the vertex pairs via weighted two-hop paths, one hop through ${\bf G}^N$ followed by one through ${\bf G}^D$, a new edge is added between any two vertices whose total path weight exceeds the threshold, \( \tau \). Similarly, the same rule applies to two-hop paths from ${\bf G}^D$ followed by ${\bf G}^N$. The sparsity pattern of the thresholded graph, ${\bf G}_\tau$, then determines the graph for the Fermi-operator expansion of the density matrix, which can be partitioned over the subgraphs.
By tuning the exponential decay, $\alpha$, and the numerical threshold, $\tau$, this provides an efficient dynamical graph adaptation with tunable accuracy.

In practice, ${\bf G}^N$ for each subgraph is first built using the neighbor list, and then multiplied by the atom-condensed density matrix ${\bf G}^D$ associated with that subgraph, followed by the application of a numerical threshold. ${\bf G}_{\tau}$ is subsequently obtained by graph collection and symmetrization, ensuring that connectivity remains consistent and complete across partition boundaries.

The graph adaptation in Eq.\ (\ref{Graph_Adaptation}) is different from previous implementations, where we combined unweighted graphs from the thresholded atom-condensed density matrix and Hamiltonian and then added paths of length two (i.e., neighbors of neighbors) \cite{ANiklasson16, CNegre_2022}. This approach provides high accuracy, but includes too many redundant edges in the halos, which significantly reduces performance.

\begin{figure}[htbp]
    \centering
    \includegraphics[width=0.49\textwidth]{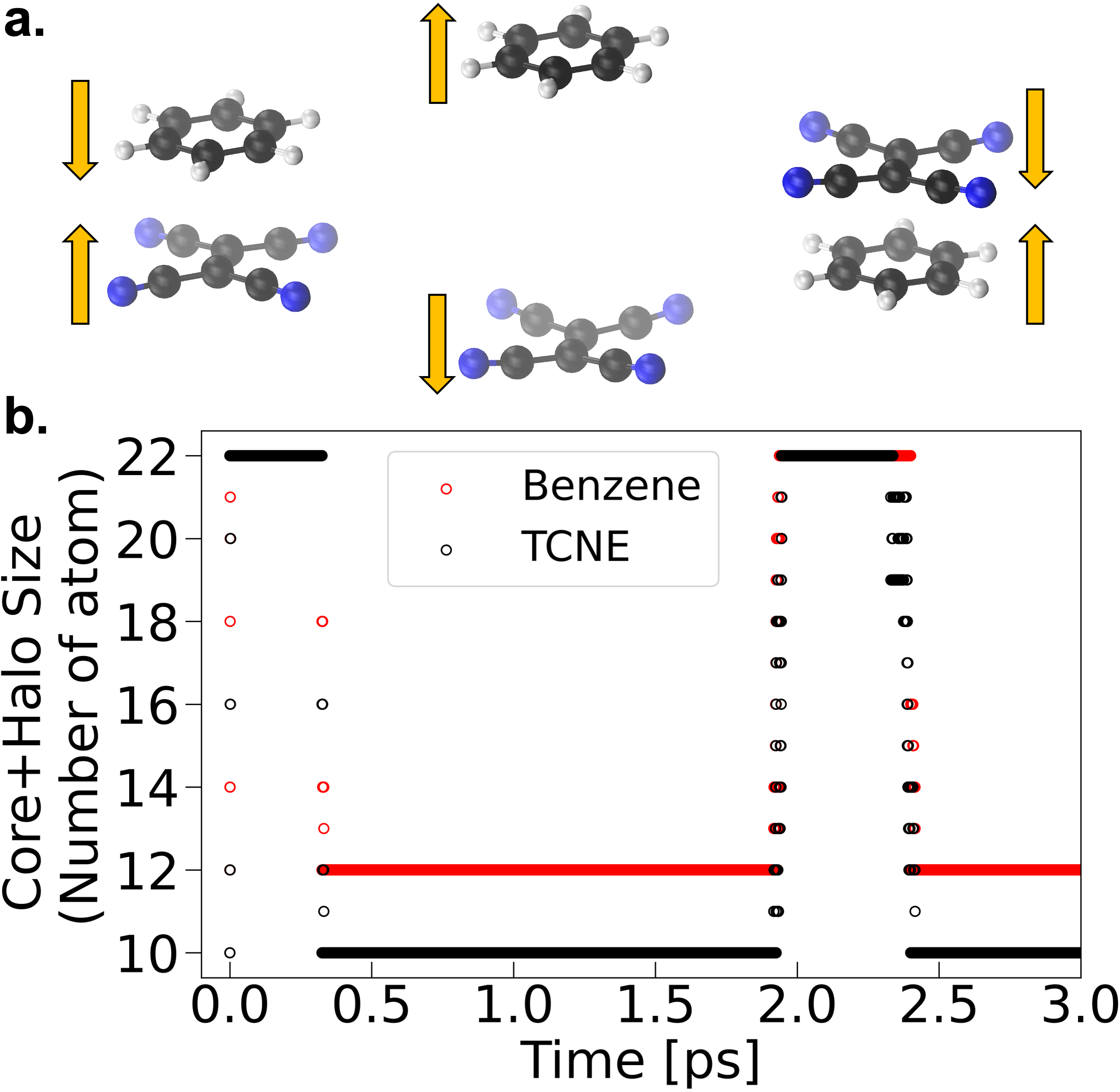}
    \caption{(a) Snapshots of the benzene–TCNE complex showing the dynamics of the two molecules approaching and separating from each other, and (b) evolution of the sizes of the two subgraphs during the MD simulation.  The integration time step is $\delta t = 0.25$~fs and the electronic temperature is set to $T_e$ = 300~K. Red and black hollow circles correspond to the size of the two subgraphs (core+halo) for the benzene and TCNE core regions, respectively.}
    \label{corehalo_fluctuation}
\end{figure}

To demonstrate that the proposed graph adaptation can dynamically expand and contract the graph while maintaining physical correctness and accuracy, we simulate a benzene–tetracyanoethylene (TCNE) complex in a periodic box, as illustrated in Fig.~\ref{corehalo_fluctuation}a. In this setup, the two molecules were initially positioned 4~\AA\ apart, with their molecular planes oriented parallel to each other. An initial velocity of about 0.002~\AA/fs was assigned to both molecules in opposite directions to ensure zero total momentum. This configuration and use of a microcanoncial (NVE, constant energy) QMD simulation (without thermostat) provide a sensitive test to gauge the new dynamic graph adaptation in SEDACS. We expect rapid changes in the intermolecular electronic connectivity as the two nearly planar fragments connect and separate during the QMD simulation, while the total energy should remain stable.

The initial graph partitioning produced two subgraphs, with the core atoms corresponding exactly to the benzene and TCNE molecules, respectively. The two initial core regions have no electronic overlap with each other and thus no halos (core only). As the QMD simulation begins, the two molecules move toward each other. As shown in Fig.~\ref{corehalo_fluctuation}b, the two subgraphs (core + halo) then quickly expand to include all 22 atoms. Subsequently, as the molecules rebound and separate, each subgraph (core + halo) contracts back to the size of its respective molecule (core only). In later stages, both molecules encounter their periodic images near the simulation boundaries, leading to another expansion of the subgraphs to include the full system. These observations demonstrate the capability of our dynamic graph adaptation scheme to capture the necessary inter-partition connectivity while efficiently eliminating redundant connections as the atomic configuration evolves throughout the simulation.

To further evaluate and demonstrate the dynamic graph adaptation in the NVE QMD simulations, we explored the behavior of the total free energy (potential + kinetic)
using three different integration time steps, $\delta t = 0.5$, 0.25, and 0.125~fs, at two electronic temperatures, $T_e = 300$ and 10,000~K. Each simulation was performed without thermostatting using the NVE ensemble, which means that the total free energy (potential + kinetic, including the electronic entropy) should be conserved.

\begin{figure}[htbp]
    \centering
    \includegraphics[width=0.49\textwidth]{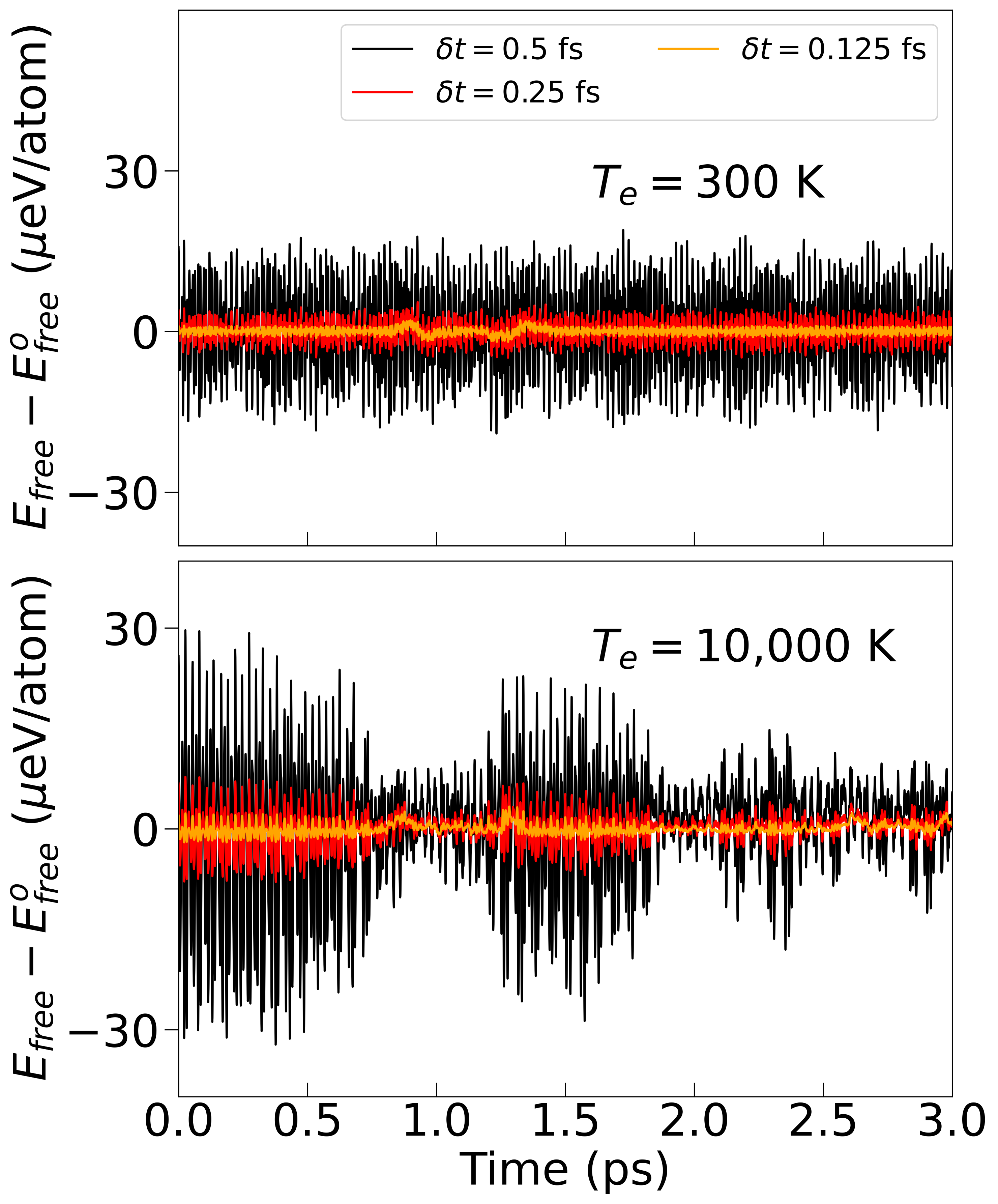}
    \caption{Total free energy fluctuations (kinetic + potential energy, including the electronic entropy contribution) from the benzene-TCNE complex simulations using three integration time steps, $\delta t = 0.5$, 0.25, and 0.125~fs, under two electronic temperatures, $T_e = 300$ and 10,000~K.}
    \label{graph_adaption_300K_10000K}
\end{figure}

In Fig.~\ref{graph_adaption_300K_10000K}, we present the total free energy fluctuations of these simulations. The figure shows that the amplitude in the total free energy fluctuations increases by approximately a factor of four when the time step is doubled, consistent with the global $O(\delta t^2)$ error of the velocity-Verlet-based integration scheme that we use \cite{ANiklasson09,PSteneteg10,GZheng11}. Furthermore, no abrupt changes or systematic long-term drift appear in the total free energy fluctuations as the subgraphs expand and contract during the QMD simulations. Together, these observations indicate that the energy functions and forces derived from the separate subgraph calculations behave as expected for the composite system and remain stable throughout the simulations. This consistency persists also at elevated electronic temperatures, where the electronic entropy contribution becomes significant, demonstrating that our method also accurately captures systems with fractional orbital occupation numbers. This is particularly important in order to provide stability and avoid degeneracies in chemically active systems. 

\subsection{Accuracy Control via Graph Thresholding}

By adjusting the graph threshold parameter, $\tau$, the size of the halo regions associated with each subgraph can be systematically controlled, achieving a balance between computational efficiency and precision. This threshold effectively serves as a tunable parameter that determines the extent of interatomic connectivity retained during the simulation. We could also choose to tune the additional $\alpha$ parameter determining the exponential decay of the weighted neighbor graph, ${\bf G}^N$. We have empirically found that using a fixed alpha parameter ($\alpha = 0.7$) works fine for all our simulations.
By varying only the graph threshold, $\tau$, we can systematically modulate the accuracy of the simulation results, enabling a controllable trade-off between physical fidelity and computational cost.

For our next test, we performed a series of SCF optimizations on a water box consisting of 2,955 atoms. As a reference, we used a single graph partition of the entire system, with the SCF convergence criterion set to $10^{-10}$ for the root-mean-square error (RMSE) of the charges. Subsequently, we carried out a series of SCF optimizations where the water box was divided into 12 partitions, employing graph thresholds ranging from $10^{-2}$ to $10^{-6}$ to systematically assess the effect of graph sparsity on accuracy.

\begin{figure}[htbp]
    \centering
    \includegraphics[width=0.49\textwidth]{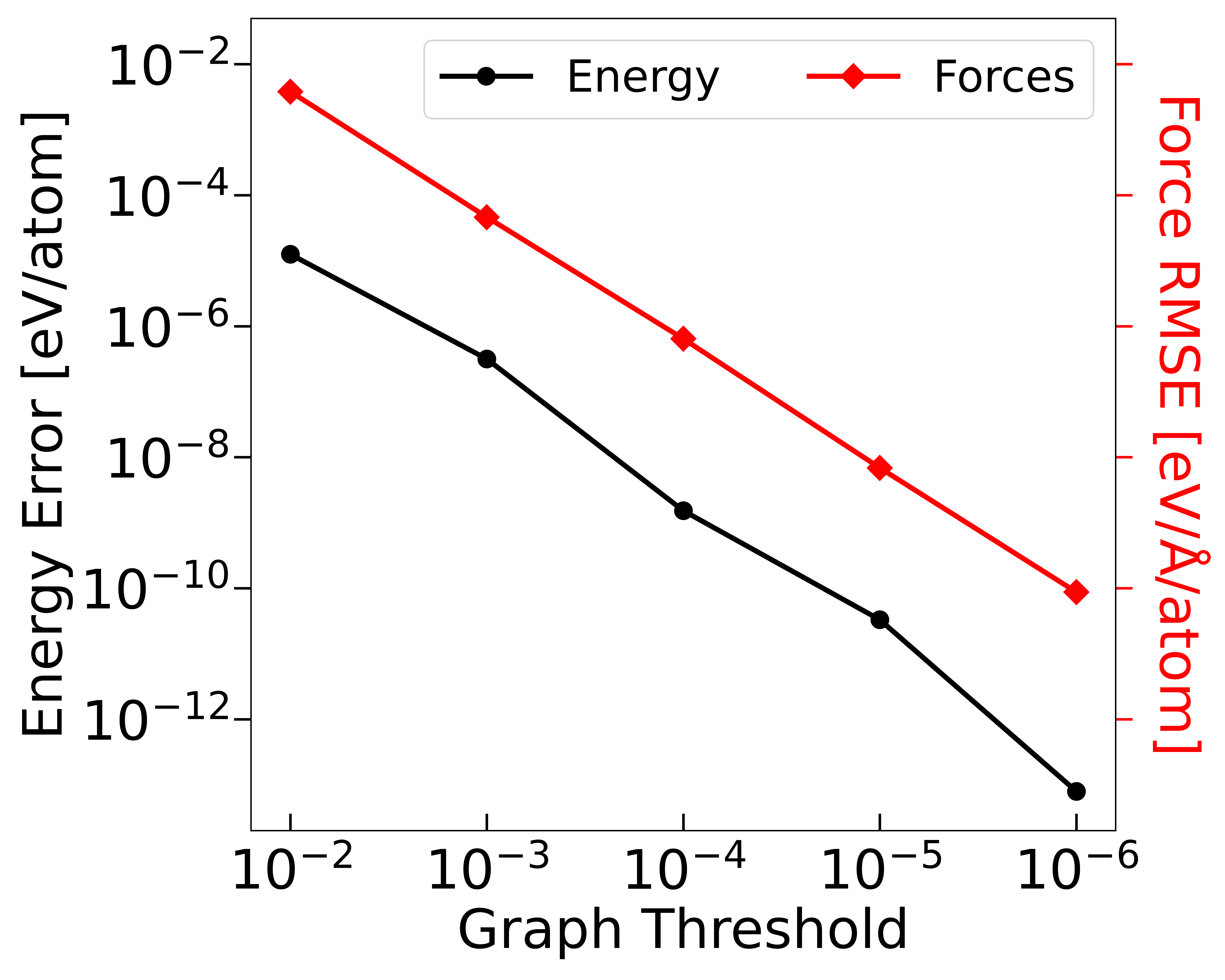}
    \caption{Errors in the free energy potential (left y-axis) and RMSE of atomic forces (right y-axis) obtained after SCF optimization for a 2,955-atom water box using different graph thresholds, $\tau$, for simulations with 12 partitions, compared to the reference simulation performed with a single partition. The decay parameter for the weighted distance graph was set to $\alpha = 0.7$.}
    \label{energy_forces_errors}
\end{figure}

The accuracy of each graph threshold was evaluated by calculating the absolute deviation in total energy and the RMSE of the atomic forces with respect to the reference single-partition simulation. In Fig.~\ref{energy_forces_errors}, the errors in total energy and atomic forces are plotted against the graph threshold on a log–log scale. Both metrics follow a near-linear relationship, suggesting a power-law dependence in which the errors decrease systematically with decreasing graph threshold. A likely explanation for the power-law behavior is the exponential decay of the electronic overlap between the core partitionings, which is naturally captured by our adaptive graph scheme. 

The decay in the force and potential-energy errors as a function of the numerical threshold demonstrates the tunable accuracy of the graph-based approach. Indirectly, it also shows that a single numerical threshold can capture the non-local electronic overlap, even though that overlap generally is non-uniform. This is consistent with a numerically thresholded sparse-matrix algebra approach, as expected, because the combined results of the partitioned subgraph calculations correspond directly to those of a sparse-matrix calculation performed on the full graph, which is determined by the numerical threshold.

\subsection{Scalability}

To assess the parallel performance of the SEDACS code across multiple processes in applications of large-scale MD simulations, we perform both strong- and weak-scaling tests. For the strong-scaling test, we ran MD simulations for a 30,840-atom water box, with 1, 2, 4, 8, and 16 GPU nodes. The entire system was partitioned into 5,120 subgraphs in all simulations. A graph threshold of $\tau = 0.001$ was used in the simulations (and with $\alpha = 0.7$), which resulted in an average halo size of about 100 atoms per partition. All scaling tests were performed on Perlmutter GPU nodes at the National Energy Research Scientific Computing (NERSC) Center. Each node was equipped with a single AMD EPYC 7763 CPU and four NVIDIA A100 GPUs and 16 MPI ranks assigned to each node.

\begin{figure}[htbp]
    \centering
    \includegraphics[width=0.49\textwidth]{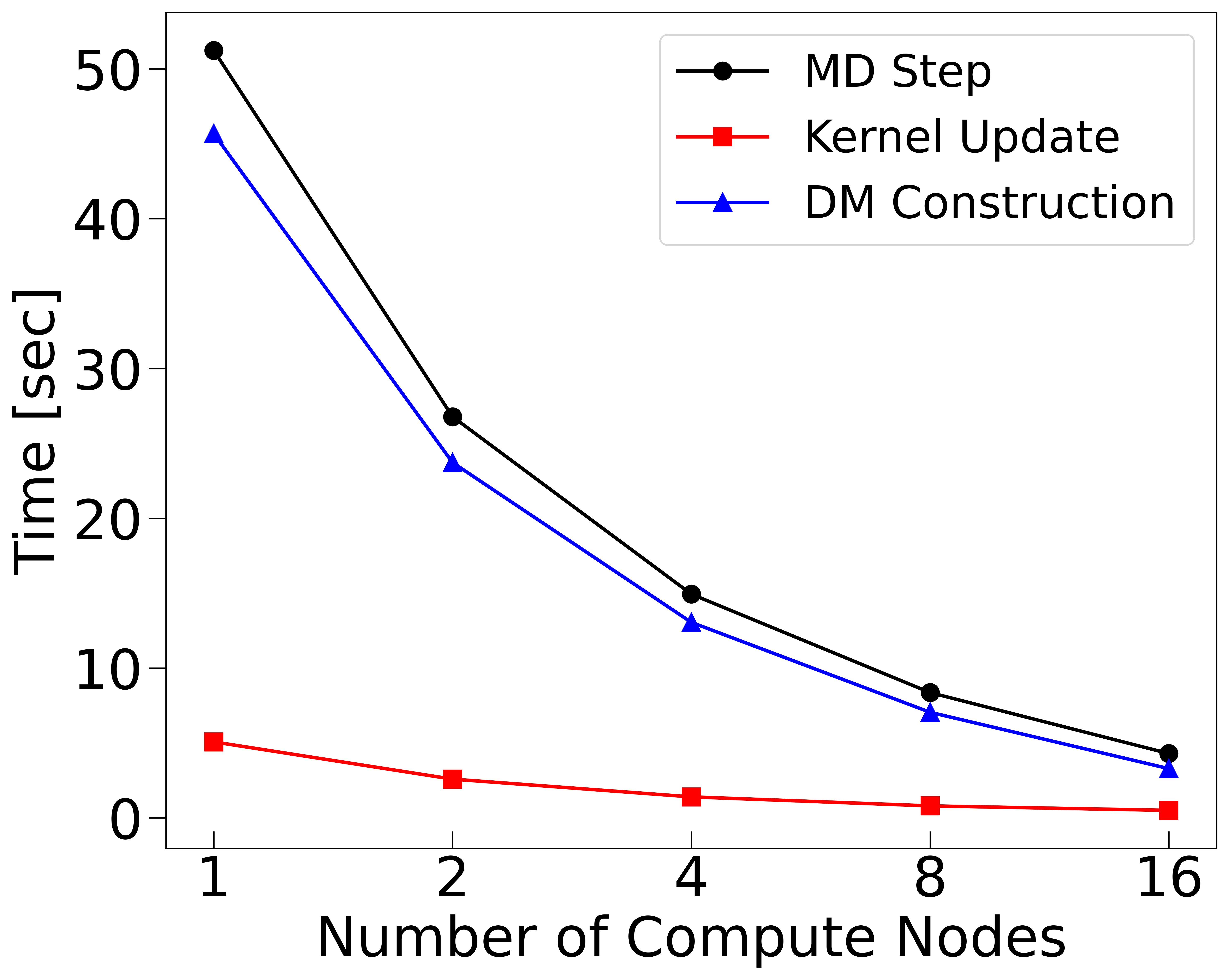}
    \caption{Strong-scaling benchmark for a 30,840-atom water box, illustrating the wall-clock time of a single MD step (black), low-rank kernel update (red), and density-matrix construction (blue) on 1, 2, 4, 8, and 16 GPU nodes.}
    \label{strong_scaling}
\end{figure}

After the initial full SCF optimization in the first time step, we collected the timings for the full MD steps (MD Step), as well as the kernel update associated with the low-rank preconditioned Krylov subspace approximation (Kernel Update), and the density-matrix (DM) construction performed through the SEDACS-LATTE interface (DM Construction). For the low-rank kernel updates, an average of three rank updates were performed using a relative residual error tolerance of $10^{-2}$. As shown in Fig.~\ref{strong_scaling}, the time required for a single MD step improves by factors of approximately 1.9, 3.4, 6.1 and 11.9 when the number of GPU nodes is increased to 2, 4, 8, and 16, respectively. In comparison, the DM construction achieves speedups of 1.9, 3.5, 6.3 and 13.9, while kernel updates show improvements of 2.0, 3.6, 6.3, and 10.0. These results indicate that the overall scaling for the MD step is primarily limited by the relatively weaker scaling of the kernel updates. This is alleviated by the Ewald solver implemented in SEDACS which utilizes GPUs to accelerate the all-to-all Ewald-based Coulomb summation over the entire periodic system \cite{MCKaymak25}. The DM construction requires the evaluation of the global chemical potential \cite{CNegre_2022} and the MPI communication of full-system data, such as graphs, charges, and forces, also require operations collecting data over the entire system, however, these operations are not the dominant bottleneck. Instead, the DM construction inefficiency arises primarily from halo-size imbalances, with some MPI ranks handling much larger matrices than others. In our test case using 16 GPU nodes, uneven halo sizes cause the most expensive MPI rank to incur a DM construction cost 1.37$\times$ higher than the least expensive, where cost is estimated as the sum of cubed halo sizes. Strategies for minimizing this load imbalance will be explored in the future.

For the weak-scaling test, we began by performing an MD simulation of a 2,088-atom water box on a single GPU node. The number of GPU nodes was then increased to 2, 4, 8, and 16 while maintaining a constant ratio between the system size and the number of GPU nodes. The system was initially partitioned into 320 subgraphs for the single-node simulation, with the number of subgraphs scaled proportionally as additional nodes were added. A graph threshold of $\tau = 0.001$ (and with $\alpha = 0.7$) was used in the simulations, which resulted in an average halo size of about 110 atoms per partition. 

\begin{figure}[htbp]
    \centering
    \includegraphics[width=0.49\textwidth]{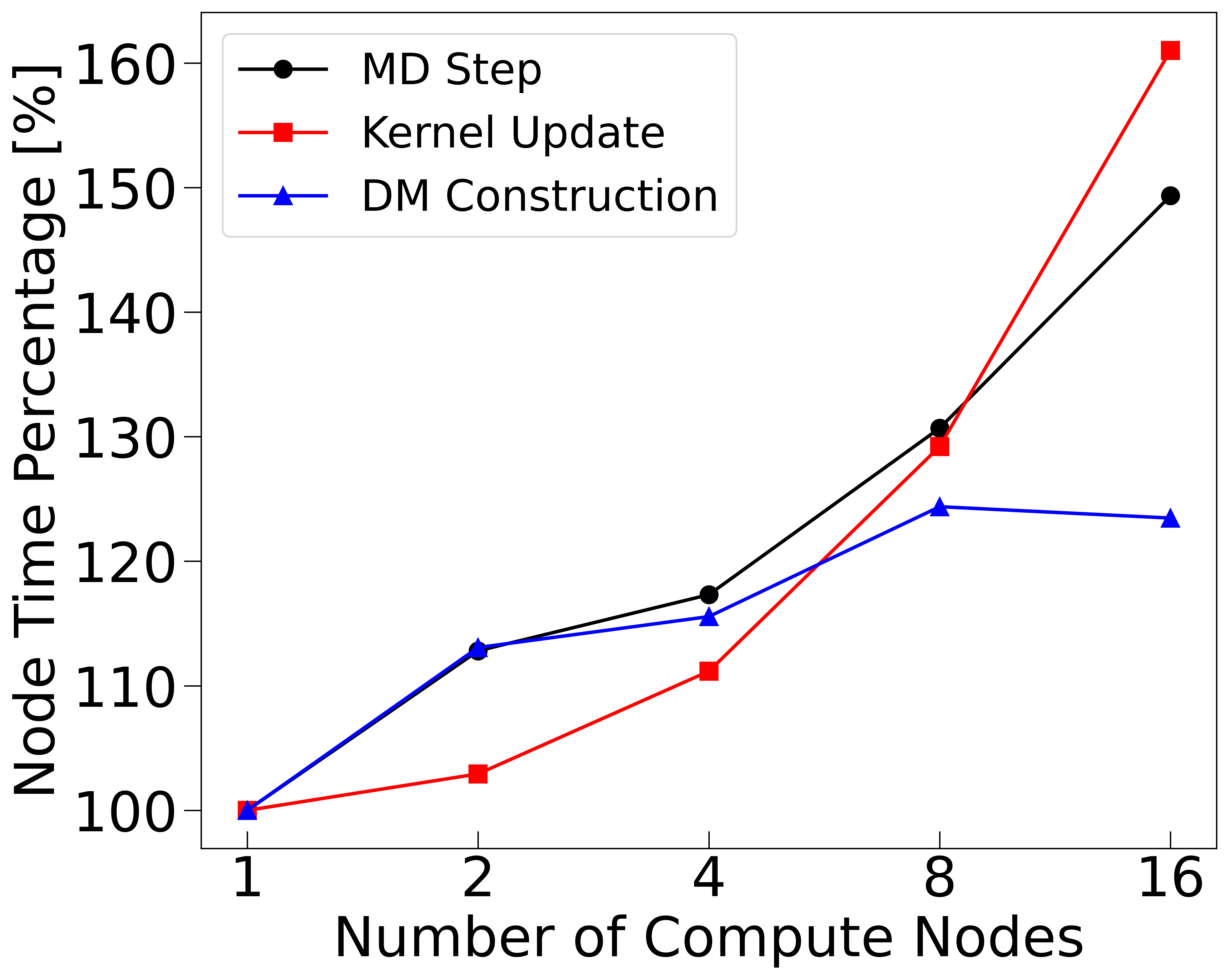}
    \caption{Weak-scaling benchmark starting with 2,088 water atoms for a single node and ending with 33,408 water atoms for the simulation on 16 nodes. The plot shows the wall-clock time of each case, normalized to the single-node runtime (in percentage), for a single MD step (black), low-rank kernel update (red), and density-matrix construction (blue) on 1, 2, 4, 8, and 16 GPU nodes.}
    \label{weak_scaling}
\end{figure}

We collected timings after the initial SCF optimization for the full MD step (MD Step), the kernel updates (Kernel Update), and the DM construction (DM Construction). For the low-rank kernel updates, an average of three rank updates were performed with a relative residual error tolerance of $10^{-2}$. As shown in Fig.~\ref{weak_scaling}, the single-node timing was used as the baseline, and subsequent timings were expressed as a percentage relative to this reference. In an ideal weak-scaling scenario, where the system size increases proportionally with the number of GPU nodes, the timing should remain constant at 100\% of the single-node runtime. At 16 GPU nodes, the wall-clock times for the MD step, kernel update, and DM construction increased by factors of approximately 1.5, 1.2, and 1.6, respectively. The observed parallel inefficiency again arises from full-system operations and largely due to load imbalance in the halo sizes. In the weak-scaling test, however, the impact of full-system operations becomes more and more significant (especially for the kernel updates, which require all-to-all Coulomb summations), because the full system size grows with the number of GPU nodes. 

\subsection{High-Explosive Graph-Based QMD Simulations}

To demonstrate the ability of the new SEDACS framework to perform scalable QMD simulations of complex, chemically active systems, we ran NVE simulations of a high-explosive RDX (1,3,5-Trinitro-1,3,5-triazinane) system, including tens-of-thousands of atoms, using a high initial temperature. The RDX system was constructed by replicating the X-ray crystal structure \cite{rdx_crystal} into a $5 \times 5 \times 5$ supercell, resulting in a total system size of 21,000 atoms and a density of 1.85 g/cm$^3$. The entire system was partitioned into 2,560 subgraphs and distributed across 16 GPU nodes (using the same hardware configuration as in the previous scaling tests). A graph threshold with $\tau = 3 \times 10^{-4}$ was used in the simulations (and $\alpha = 0.7$), resulting in an average halo size of about 370 atoms per partition. The SEDACS simulations were performed at an electronic temperature of 2,500~K, and the system was initialized with velocities sampled from a Maxwell–Boltzmann distribution at the same temperature along with uniformly sampled displacements of atomic coordinates from 0 to 1.3~\AA. The low-rank kernel updates required an average of seven rank updates using a relative residual error tolerance of 10$^{-2}$. 

\begin{figure}[t]
    \centering
    \includegraphics[width=0.49\textwidth]{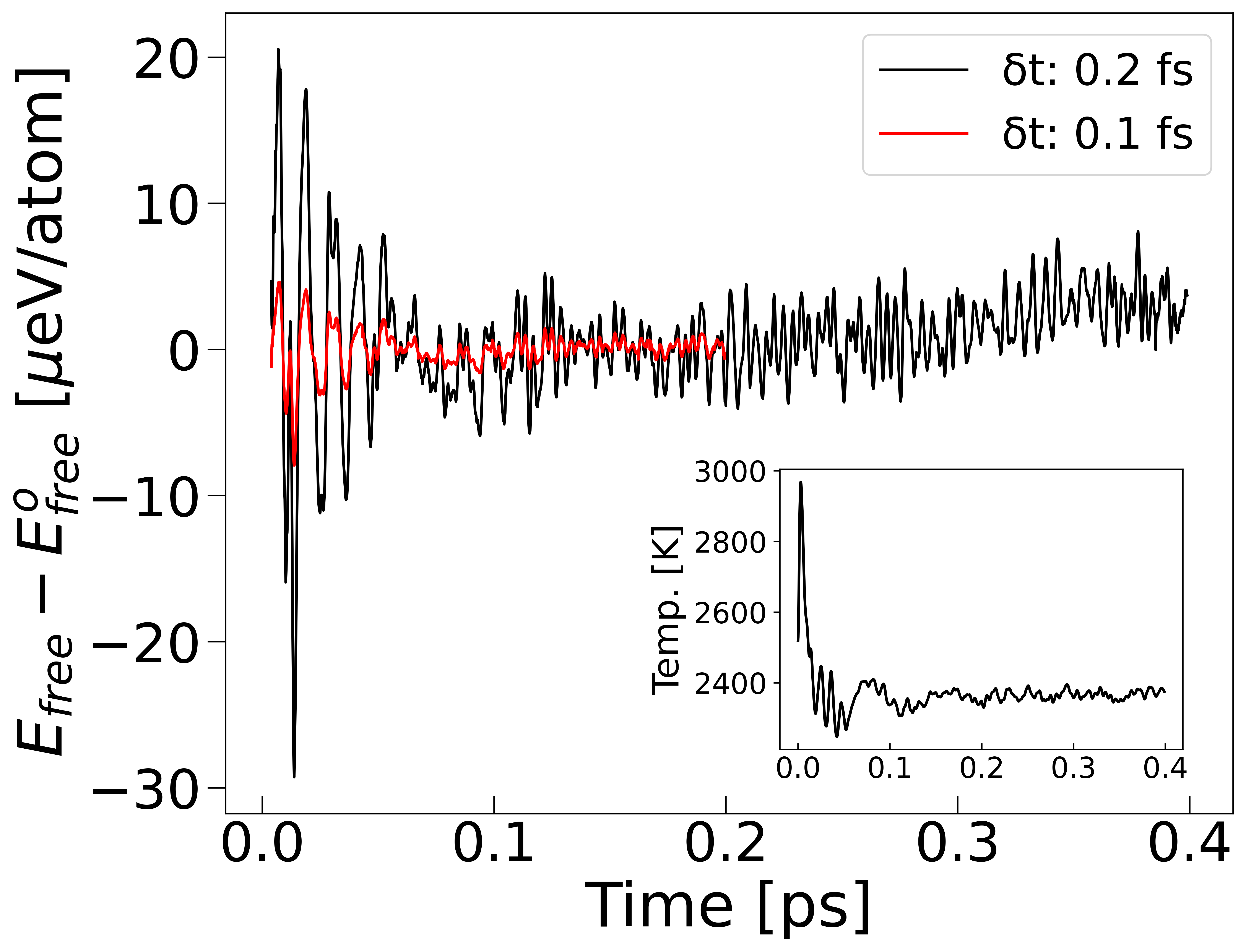}
    \caption{Total free energy fluctuation (kinetic + potential energy, including the electronic entropy contribution) from NVE simulations  of the 21,000-atom RDX system at a density of 1.85 g/cm$^3$ using two integration time steps, $\delta$t = 0.2 and 0.1~fs, over 2,500 MD steps. The inset figure shows the statistical temperature for the simulation with $\delta$t = 0.2~fs.}
    \label{rdx}
\end{figure}

The RDX simulations were run with SEDACS for 2,000 time steps in the NVE ensemble using two different integration time steps, $\delta t = 0.2$ and 0.1~fs. In Fig.~\ref{rdx}, the total free energy fluctuations are well-behaved and stable. As expected from the second-order velocity-Verlet-based integration scheme, doubling the time step from 0.1 to 0.2~fs leads to an approximately fourfold increase in the magnitude of the energy fluctuations. Thermal decompositions such as NO$_2$ release \cite{AStrachan03, JYWu23} were observed in the QMD simulations, indicating the onset of chemical reactivity at elevated temperature. Despite these highly reactive conditions, the graph-based QMD simulations remained numerically stable and scalable to the size of tens-of-thousands of atoms.

\section{Summary and Conclusions}

In this work, we introduced the Scalable Ecosystem, Driver, and Analyzer for Complex Chemistry Simulations (SEDACS), a modular and extensible Python-based framework that enables graph-based linear-scaling QMD simulations with external electronic-structure codes. 
Requiring only minor code modifications, SEDACS makes it possible to perform parallel simulations using graph-based linear-scaling electronic-structure theory and XL-BOMD with shadow potentials, achieving both controlled accuracy and substantial gains in speed and scalability.

We showed that SEDACS provides rigorous and tunable error control through graph thresholding, allowing systematic trade-offs between computational cost and accuracy in energies and forces. The dynamic graph adaptation strategy further improves efficiency by maintaining small halo sizes while preserving accuracy and stability as atomic configurations and electronic connectivity evolve during simulations. These SEDACS capabilities were  analyzed using an external SCC-DFTB-based Fortran code in applications involving large, complex and chemically reactive systems with tens-of-thousands of atoms, demonstrating scalability and stable dynamics even in charge-sensitive chemically active environments.

By combining graph-based linear-scaling electronic-structure theory with shadow MD, SEDACS eliminates the need for iterative SCF convergence at every MD step, while maintaining rigorous error control and long-term energy stability. The incorporation of a low-rank Krylov subspace approximation for kernel updates enables efficient propagation of electronic degrees of freedom and accelerates both QMD simulations and the initial SCF optimization. 

Most notably, the SEDACS ecosystem enables stable QMD simulations using external electronic-structure codes of chemically active systems containing tens-of-thousands of atoms, a regime computationally inaccessible to the external solvers alone. These SEDACS-driven simulations remain stable even in the presence of bond breaking and formation, underscoring the tunable robustness of the graph-based QMD approach within the shadow molecular dynamics framework.

This work firmly establishes the utility of SEDACS as a framework for bringing graph-based linear-scaling electronic-structure theory and shadow MD to existing electronic-structure codes. By decoupling algorithmic innovation from the external electronic-structure code development, SEDACS lowers the barrier to large-scale first-principles simulations and opens the door to future extensions, including new electronic-structure methods, heterogeneous hardware architectures, and data-driven acceleration strategies.


\section{Data Availability}

The graph-enabled SCC-DFTB based LATTE  code and SEDACS are publicly available at \url{https://github.com/lanl/sedacs}. Standalone LATTE is available at \url{https://github.com/lanl/LATTE} along with documentation and tutorials.

\section{Acknowledgements}

This work is supported by the U.S. Department of Energy Office of Basic Energy Sciences (FWP LANLE8AN), the Los Alamos National Laboratory LDRD program and by the U.S. Department of Energy through the Los Alamos National Laboratory. This research used resources provided by the Darwin testbed at Los Alamos National Laboratory (LANL) which is funded by the Computational Systems and Software Environments subprogram of LANL's Advanced Simulation and Computing program (NNSA/DOE). Additionally, this research used resources of the National Energy Research Scientific Computing Center (NERSC), a Department of Energy User Facility using NERSC award BES-ERCAP 37311. The authors also acknowledge the computing resources provided by LANL Institutional Computing. Los Alamos National Laboratory is operated by Triad National Security, LLC, for the National Nuclear Security Administration of the U.S. Department of Energy Contract No. 892333218NCA000001. This article has been approved for unlimited distribution with the LA-UR number: LA-UR-26-23784

\bibliography{ref}
\cleardoublepage

\section*{Supporting Information}
\subsection{Matrix Polynomials on a Graph}

In matrix algebra, a sparse matrix, $\mathbf{X}$, can be expressed as a graph, $\mathbf{G}$, where the vertices represent the diagonal matrix elements and the edges represent the non-zero off-diagonal elements connecting the vertices.

A sparse matrix polynomial (or analytic function), $f(\mathbf{X}) = \sum_k c_k {\mathbf{X}^k}$, can be be evaluated, step-by-step, through successive matrix multiplications and additions/subtractions, while discarding elements below some chosen global numerical threshold, $\tau$, at each step of the polynomial expansion. By the step-by-step expansion, we mean that the polynomial order of ${\mathbf{X}^k}$ increases only in steps of 1 with matrix multiplications applied only from the left (or right). By a global threshold, we mean that any matrix element, $X_{ij}$, whose magnitude exceeds $\tau$ at any point during the expansion, is kept for the whole computation (e.g.\ see Ref.\  \cite{CNegre_2022} and its supporting information for more details). The global threshold is different from a more conventional numerically thresholded sparse matrix algebra, where we typically remove matrix elements smaller than some threshold after each matrix-matrix operation. The incremental step-by-step expansion is also different from some more optimized expansions schemes where, for example, we may calculate ${\mathbf{X}^4}$ as ${\mathbf{X}}^2 \times {\mathbf{X}}^2$, ignoring the intermediate ${\mathbf{X}^3}$. The non-zero matrix elements, $|X_{ij}|$, that at any point appear above the threshold $\tau$ can be represented by a graph, $\mathbf{G}_{\tau}$. We will refer to the globally thresholded step-by-step sparse matrix expansion defined above as a \textit{matrix polynomial on a graph}, denoted by $f_G(\mathbf{X})$ or $f(\mathbf{X})\big \vert_G$. 

The graph $\mathbf{G}_{\tau}$ (or simply $\mathbf{G}$) can be partitioned into smaller overlapping subgraphs, $\{{\mathbf g}_i\}$, consisting of non-overlapping core parts (c) and overlapping halos. For example, each vertex could be chosen as a core with all its edges and nearest neighboring vertices belonging to the halo. These subgraphs can then be used to extract the corresponding principal submatrices, $\{\mathbf{x}_i\}$, of $\mathbf{X}$ \cite{ANiklasson16}.

The central observation behind graph-based linear-scaling electronic structure theory is the equivalence between a sparse matrix function on a graph $\mathbf{G}$ (or the equivalent globally thresholded step-by-step matrix function expansion) and a collection of the core parts of the dense matrix functions over the principal submatrices of $\mathbf{X}$, i.e., \
\begin{align}
f_G(\textbf{X}) &  = \Bigl\{f_c(\textbf{x}^{(i)})\Bigr\}_{\text{collect}}.
\end{align}
This relation means that there is a one-to-one mapping between a numerically thresholded sparse matrix algebra and a divide-and-conquer like approach, providing a natural way to parallelize calculations while controlling the error with an adjustable numerical threshold.

\subsection{Low-Rank Kernel Approximation}
Using matrix--vector notation, the kernel defined in the equation of motion for the electronic degree of freedom can be written as
\begin{align}
    \mathbf{K} = \mathbf{J}^{-1},
    \ J_{ij} = \frac{\partial f_i(\mathbf{n})}{\partial n_j}, 
    \ \mathbf{f} = \boldsymbol{\rho}_0[\mathbf{n}] - \mathbf{n}, 
    \label{Jacobian}
\end{align}
where \( \mathbf{K}, \mathbf{J} \in \mathbb{R}^{N \times N} \) and 
\( \mathbf{f}, \mathbf{n}, \boldsymbol{\rho}_0[\mathbf{n}] \in \mathbb{R}^{N} \).

With these definitions, the electronic equation of motion can be rewritten in the equivalent form
\begin{align}
    \ddot{\mathbf{n}} 
    = - \omega^2 \left( \mathbf{K}_0 \mathbf{J} \right)^{-1} 
    \mathbf{K}_0 \left( \boldsymbol{\rho}_0[\mathbf{n}] - \mathbf{n} \right),
    \label{EEOM}
\end{align}
where \( \mathbf{K}_0 \) denotes an approximate inverse of the Jacobian, 
\( \mathbf{K}_0 \approx \mathbf{J}^{-1} \). By construction, this implies 
\( (\mathbf{K}_0 \mathbf{J})^{-1} \approx \mathbf{I} \). The central idea is that now the action of this operator on the preconditioned residual vector \( \mathbf{K}_0 \mathbf{f} \) can be accurately represented using a low-rank approximation.

In the exact formulation, the operator \( (\mathbf{K}_0 \mathbf{J})^{-1} \) can be expressed through the generalized decomposition
\begin{align}
    (\mathbf{K}_0 \mathbf{J})^{-1}
    = \sum_{i,j=1}^{N} \mathbf{v}_i \, M_{ij} \, \mathbf{f}_{\mathbf{v}_j}^{T},
    \label{PreKernel}
\end{align}
where
\begin{align}
    \mathbf{f}_{\mathbf{v}_j}
    \equiv 
    \mathbf{K}_0 \left.
    \frac{\partial \mathbf{f}(\mathbf{n} + \lambda \mathbf{v}_j)}{\partial \lambda}
    \right|_{\lambda = 0}
    = \mathbf{K}_0 \mathbf{J} \mathbf{v}_j,
    \label{dfdv}
\end{align}
for an arbitrary complete set of vectors \( \{ \mathbf{v}_j \} \). The vectors
\( \{ \mathbf{f}_{\mathbf{v}_j} \} \) thus represent the corresponding set of
preconditioned directional derivatives. The matrix elements \( \{ M_{ij} \} \)
are defined through
\begin{align}
    \mathbf{M} = \mathbf{O}^{-1},
    \qquad
    O_{ij} = \mathbf{f}_{\mathbf{v}_i}^{T} \mathbf{f}_{\mathbf{v}_j}.
\end{align}

In practice, a low-rank approximation of \( (\mathbf{K}_0 \mathbf{J})^{-1} \) in
Eq.~(\ref{PreKernel}), acting on the preconditioned residual
\( \mathbf{K}_0 \mathbf{f}(\mathbf{n}) \) in Eq.~(\ref{EEOM}), can be constructed
using a reduced set of \( m < N \) carefully selected vectors
\( \{ \mathbf{v}_j \}_{j=1}^{m} \).
These vectors may be chosen from an orthogonalized preconditioned Krylov
subspace,
\begin{align}
    \{ \mathbf{v}_j \}
    &= \mathrm{span}^{\perp}
    \Bigl\{
        \mathbf{K}_0 \mathbf{f}(\mathbf{n}), \nonumber\ (\mathbf{K}_0 \mathbf{J}) \mathbf{K}_0 \mathbf{f}(\mathbf{n}), \nonumber \\
    &\quad (\mathbf{K}_0 \mathbf{J})^{2} \mathbf{K}_0 \mathbf{f}(\mathbf{n}), \nonumber \ (\mathbf{K}_0 \mathbf{J})^{3} \mathbf{K}_0 \mathbf{f}(\mathbf{n}), \ldots
    \Bigr\}.
\end{align}

Within this preconditioned orthonormalized Krylov subspace (or Arnoldi subspace) approximation, the electronic
equations of motion can be written in the low-rank form
\begin{align}
    \ddot{\mathbf{n}}
    \approx
    - \omega^{2}
    \left(
        \sum_{i,j=1}^{m < N}
        \mathbf{v}_i \, M_{ij} \, \mathbf{f}_{\mathbf{v}_j}^{T}
    \right)
    \mathbf{K}_0
    \left(
        \boldsymbol{\rho}_{0}[\mathbf{n}] - \mathbf{n}
    \right),
    \label{LowRankEEOM}
\end{align}
which replaces the full inverse operator in Eq.~(\ref{EEOM}) by a preconditioned low-rank representation.

The dominant computational cost associated with this preconditioned kernel
approximation~\cite{ANiklasson20} arises from the evaluation of
the response vectors, \( \{\mathbf{f}_{\mathbf{v}_j}\} \), needed to construct the preconditioned Krylov subspace.
These response vectors can be computed using canonical quantum perturbation
theory with fractional orbital occupations~\cite{ANiklasson15,YNishimoto17,CNegre_2022},
which is particularly important for small-gap systems exhibiting sensitive or
unstable charge distributions.
For QMD simulations of non-reactive systems, however,
low-rank updates of \( \mathbf{f}_{\mathbf{v}_j} \) are often unnecessary.
In such cases, a fixed preconditioner \( \mathbf{K}_0 \) alone (corresponding to
\( m = 0 \)) provides sufficient accuracy.
For molecular systems with a large HOMO--LUMO gap, even a simple scaled
identity approximation,
\( \mathbf{K} \approx \mathbf{K}_0 = -c \mathbf{I} \) with \( c \in [0,1] \),
is typically adequate.
Alternatively, a preconditioner may be constructed from a regularized solution
of an approximate reference system, such as the molecular configuration at the
initial time step.
Although the construction of such a preconditioner may be computationally
expensive, reusing it over many time steps (e.g., thousands) reduces the
amortized cost to a small fraction of the total simulation effort, resulting in a net gain in overall efficiency. Details of how the preconditioner and response vectors are constructed on a graph are given in Ref.\ \cite{CNegre_2022}.

\subsection{Kernel-Accelerated SCF Optimization}

In the first time step of a shadow MD simulation we need to find a sufficiently converged SCF solution to initialize the extended dynamical electronic degrees of freedom, $n({\bf r})$, that is propagated through the harmonic oscillator equation in Eq.\ (\ref{EEOM}).
There are numerous methods to accelerate the SCF convergence. Nevertheless, for large complex and chemically active systems, SCF convergence can be a major challenge. However, the low-rank Krylov subspace technique that we use to approximate the kernel in the integration of the harmonic oscillator equations of motion can be used also in an efficient SCF-acceleration scheme. This can be seen by expressing the SCF problem in terms of finding the zero of the residual function, $\rho_0[n]({\bf r}) - n({\bf r})$, in Eq.\ (\ref{EEOM}), which  can be represented in vector form as $(\text{q}_{0}[\textbf{n}] - \textbf{n}) \in \mathbb{R}^N$. 
The kernel is defined as the inverse Jacobian of the residual function. In an iterative Newton-based SCF scheme we can therefore find the solution (using some sufficiently accurate initial guess, ${\bf n}_0$,) from 
\begin{equation*}
    {\bf n}_{k+1} = {\bf n}_k - {\bf K}\big({\bf q}_0[{\bf n}_k] - {\bf n}_k\big),
\end{equation*}
where ${\bf K} \in \mathbb{R}^{N \times N}$ is the corresponding matrix representation of the kernel, $K({\bf r},{\bf r'})$, in the equation of motion for the electronic degree of freedom. Using a preconditioner, ${\bf K}_0 \approx {\bf J}^{-1}$, which approximates the inverse Jacobian of the residual function, we can rewrite the Newton scheme as
\begin{equation*}
    {\bf n}_{k+1} = {\bf n}_k - ({\bf K}_0{\bf J})^{-1}{\bf K}_0\big({\bf q}_0[{\bf n}_k] - {\bf n}_k\big),
\end{equation*}
where the kernel has been replaced by $({\bf K}_0{\bf J})^{-1}$ acting on the preconditioned residual, ${\bf K}_0\big({\bf q}_0[{\bf n}_k] - {\bf n}_k\big)$. In this quasi-Newton scheme the Jacobian is thus approximated with a low-rank preconditioned Krylov subspace expansion of ${\bf K}_0{\bf J}$, which never has to be calculated explicitly, i.e.\ the same technique as is used to approximate the kernel in the equations of motion, Eq.\ (\ref{EEOM}). This also illustrates how the extended Lagrangian dynamics works, where the approximate extended electronic degrees of freedom, \(n({\bf r})\), are ``pushed'' dynamically toward SCF convergence along its propagation in each MD integration time step of the harmonic oscillator.
Details of the kernel approximation are given in the Section B and in Ref.\ \cite{CNegre_2022}.

To demonstrate the quasi-Newton scheme above we performed optimizations on a 30,840-atom water box and compared the convergence behavior with the conventional Pulay-scheme based on direct inversion in the iterative subspace (DIIS) \cite{PPulay80,PPulay82}. The comparisons were performed at the same two electronic temperatures as before, 300~K and 10,000~K, with the latter case involving significant fractional occupations of electronic states and therefore presenting a slightly different convergence scenario. The low-rank updates are applied independently to each subgraph using quantum-response calculations in parallel, in combination with a block-diagonal preconditioner (see \cite{CNegre_2022} for further details). On average, three rank updates are required to achieve a relative residual error tolerance of $10^{-2}$ in each SCF iteration.

\begin{figure}[htbp]
    \centering
    \includegraphics[width=\linewidth]{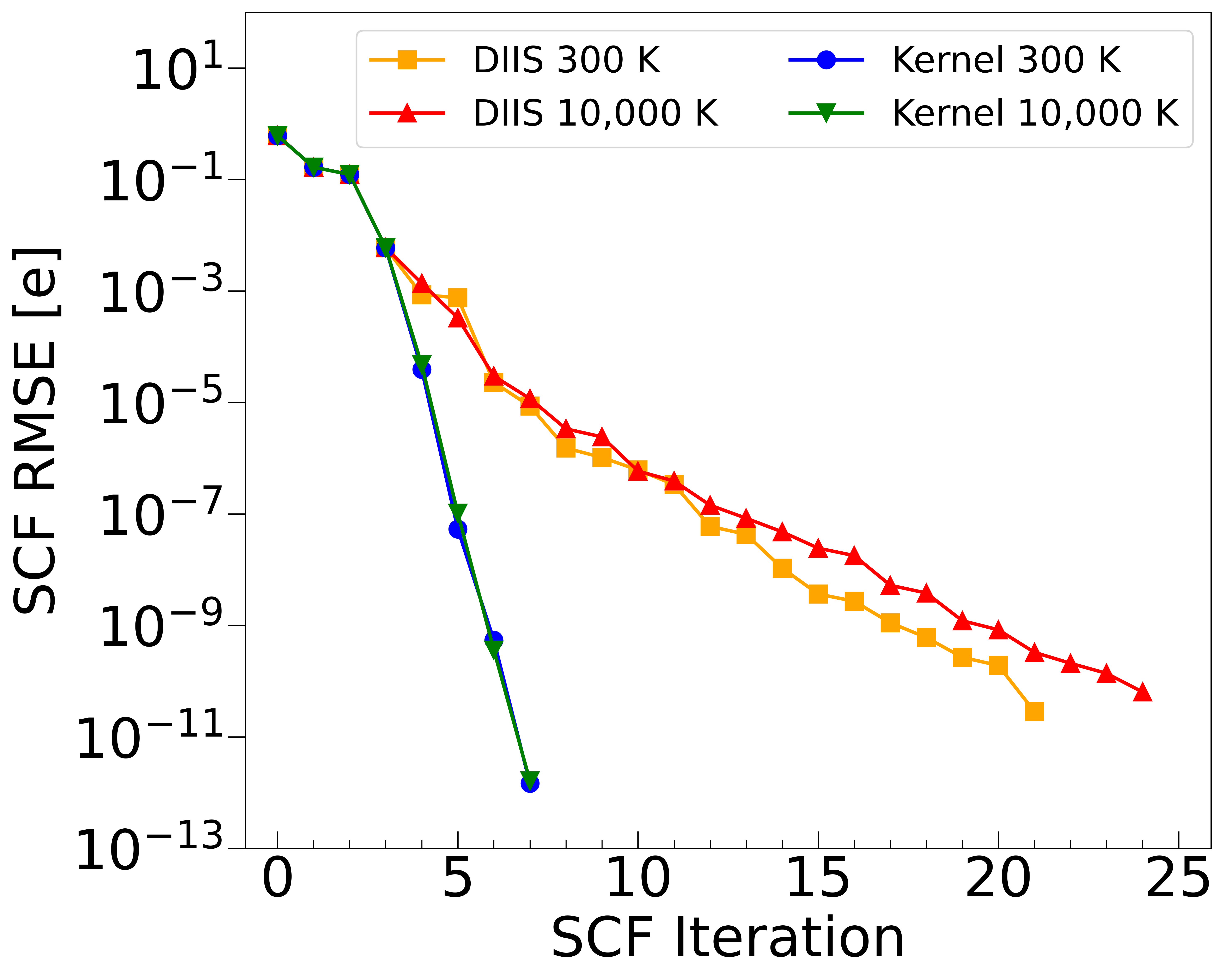}
    \caption{Comparison of SCF convergence for a 30,840-atom water box using DIIS and quasi-Newton methods at 300 and 10,000 K, quantified by the RMSE of residual charges over SCF iterations. The quasi-Newton method is started once the RMSE of the residual charges falls below 0.1.}
    \label{scf_comparison}
\end{figure}

As shown in Fig.~\ref{scf_comparison}, the quasi-Newton scheme demonstrates faster convergence compared to the DIIS method, as a function of the number of SCF iterations, at both electronic temperatures. Notably, while the DIIS approach requires a few additional iterations to reach a tight converge at 10,000~K, the quasi-Newton method converges in the same number of iterations as observed at 300~K. 

Even if we find a more rapid rate of convergence for the kernel-based quasi-Newton method, it may not always provide convergence in shorter wall-clock time compared to DIIS, because the low-rank updates increase the cost per SCF iteration. However, the kernel-based quasi-Newton method provides a robust competitive alternative and can often handle particularly difficult convergence problems. This can be of significant importance when we handle large-scale, chemically active, complex systems.


\subsection{Implementation with SCC-DFTB theory}
All simulations reported in this work were performed using an external electronic structure code based on semiempirical self-consistent-charge density-functional tight-binding (SCC-DFTB) theory~\cite{DPorezag95,GSeifert96,MElstner98,MFinnis98,TFrauenheim00,MGaus11,BAradi07,BHourahine20}, as implemented in the open-source electronic structure package \textsc{LATTE}~\cite{LATTE,AKrishnapriyan17}, together with the \textsc{PROGRESS} and
\textsc{BML} libraries~\cite{2016progress,BML}. No new parameterizations or modifications of the SCC-DFTB energy functional (apart from its shadow potential formulation) were introduced in this study.

SCC-DFTB is an approximate formulation of first-principles Kohn-Sham density functional theory derived from a second- or third-order expansion of the Kohn-Sham energy functional with respect to electronic charge fluctuations around a reference superposition of overlapping atomic densities. The method employs a minimal numerical basis set, with Hamiltonian matrix elements and overlap integrals tabulated and parameterized using the Slater-Koster formalism. Electrostatic interactions are described by screened Coulomb interactions between atomic net Mulliken partial charges. At large separations, the Coulomb interaction between atoms decays as
\( |\mathbf{R}_I - \mathbf{R}_J|^{-1} \), while at short distances it is screened to account for interactions between overlapping Slater-type charge densities. The on-site interaction term is determined by the chemical hardness or, equivalently, a Hubbard-\( U \) parameter.

Within this formalism, the charge density \( \mathbf{n} \) and the
corresponding ground-state density \( \boldsymbol{\rho}_0[\mathbf{n}] \) are represented as vectors whose components correspond to the net Mulliken partial electron occupations of each atom.

\end{document}